\documentclass[aps,prl,twocolumn,showpacs]{revtex4-1}
\usepackage{amsmath,amssymb}
\usepackage{graphicx}
\usepackage{lmodern}
\usepackage{textcomp}
\usepackage[T1]{fontenc}

\begin{document}
\title{Probing the nodal structure of Landau level wave functions in real space}
\author{J. R. Bindel$^1$, J. Ulrich$^2$, M. Liebmann$^1$, and M. Morgenstern$^1$}
\email[]{mmorgens@physik.rwth-aachen.de}
\affiliation{$^1$ II. Institute of Physics B and JARA-FIT, RWTH Aachen University, D-52074 Aachen, Germany\\
$^2$ Institute for Quantum Information and JARA-FIT, RWTH Aachen University, D-52074 Aachen, Germany}

\date{\today}
\begin{abstract}
The inversion layer of p-InSb(110) obtained by Cs adsorption of 1.8 \% of a monolayer is used to probe the Landau level wave functions within smooth potential valleys by scanning tunnelling spectroscopy at 14 T. The nodal structure becomes apparent as a double peak structure of each spin polarized first Landau level, while the zeroth Landau level exhibits a single peak per spin level only. The real space data show single rings of the valley-confined drift states for the zeroth Landau level and double rings for the first Landau level. The result is reproduced by a recursive Green's function algorithm using the potential landscape obtained experimentally.
We show that the result is generic by comparing the local density of states from the Green's function algorithm with results from a well controlled analytic model based on the guiding center approach.
\end{abstract}
\maketitle
\begin{figure*}[tbh]
\includegraphics[width=17cm]{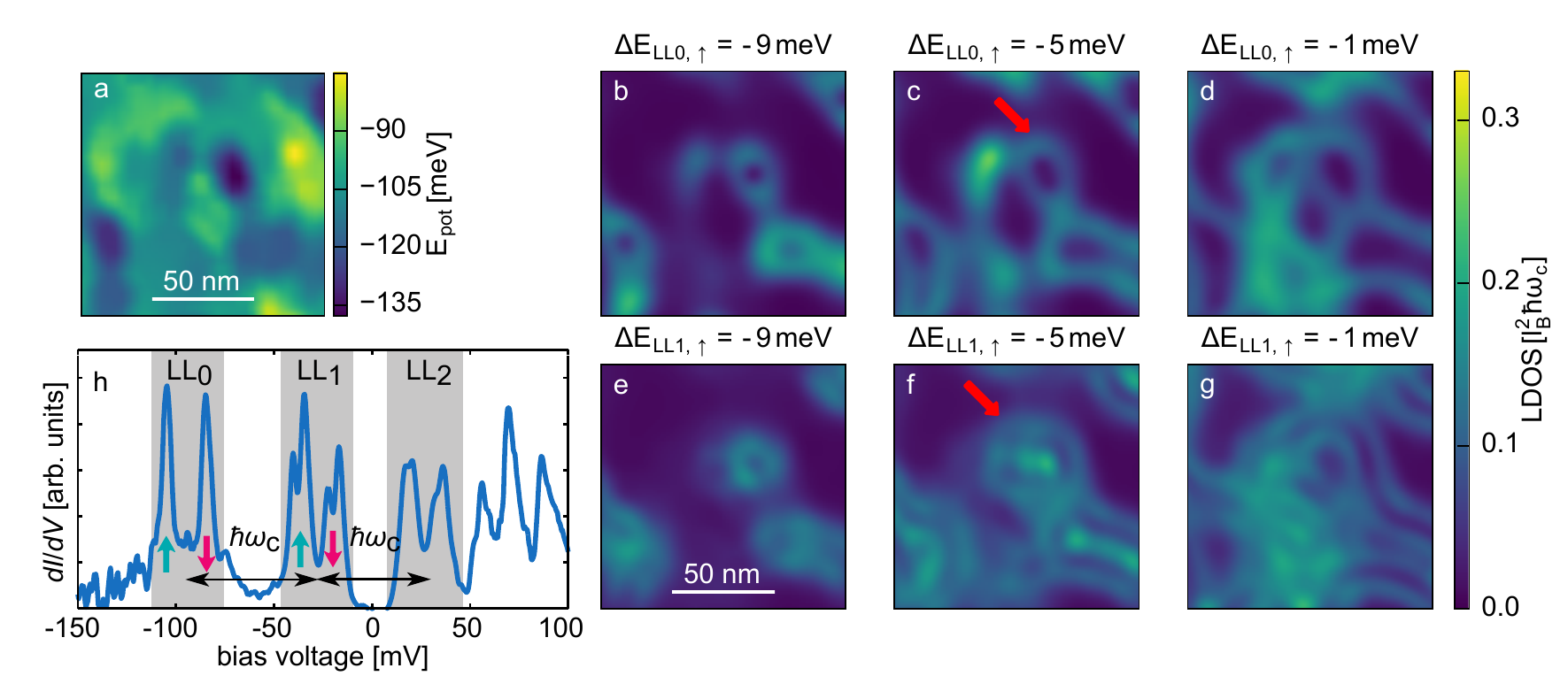}
\caption{(color online).
(a) Disorder potential of 2DES as determined by the average energy of the two spin components of LL$_0$  \cite{Bindel};
(b)-(g) LDOS within the spin level $\uparrow$ of LL$_0$ and LL$_1$; calculated by recursive Green's function algorithm \cite{recursive} using $E_{\rm pot} ({\bf r})$ of a, $m^*=0.03\cdot m_{\rm e}$, $g=-21$, $\alpha_{\rm R} =1$~eV$\rm\AA$, $B=14$~T, smooth potential boundary conditions ($>4l_B$) outside the displayed area; energies with respect to the LL center, determined by spatially averaging the LDOS, are marked on top; red arrows mark potential valley where distinct doubling of lines is observed; (h) single $dI/dV$ curve recorded within a potential valley; LL$_n$, $\hbar \omega_{\rm c}$ and opposite spin levels are highlighted, $B=14$~T, $V_{\text{stab}} = 50~\text{mV}$, $I_{\text{stab}} =150~\text{pA}$, $V_{\text{mod}} = 0.75~\text{mV}_{\text{rms}}$; differences between the two $\hbar \omega_{c}$'s relate to the known non-parabolicity of the InSb conduction band \cite{Merkt}; note the two double peaks in LL$_1$.
\label{Fig1}}
\end{figure*}
Electron wave functions are at the heart of quantum mechanics representing the particle-wave duality and determining a multitude of physical properties in solids including topologically protected  transport \cite{Zhang}. They have been probed in free space by interference experiments \cite{interference} and in solids by scanning tunneling spectroscopy (STS) of, e.g., metals \cite{Crommie}, semiconductors \cite{Morgenstern}, graphene \cite{Stroscio}, or topological insulators \cite{Yazdani}. However, one of the most fundamental single-particle wave functions, the one of quasi-free electrons within a homogeneous magnetic field, often being part of the curriculum in quantum mechanics, has never been probed directly \cite{Schine}. Without potential disorder, these wave functions are highly degenerate and not localized as described by the famous, energetically equidistant Landau levels (LLs) \cite{Landau}. However, potential disorder leads to localization of these wave functions along equipotential lines such that they can be probed in real space \cite{Ando,Morgenstern2,Hashimoto}.
Wave functions of different LLs
are distinct by their nodal structure, which is the subject of this letter \cite{Landau}. Interestingly, such nodal structure can not be probed for Dirac-type fermions, since the structure is blurred by the two-component nature of  the wave functions
\cite{Hanaguri, Andrei}.\\
Here, we probe the two antinodes of the first LL (LL$_1$) with respect to the one antinode of the zeroth LL (LL$_0$) by STS using smooth potential valleys of the quasi-free electron system of an InSb inversion layer for localization
\cite{Morgenstern3,Bindel}.  We demonstrate that the two antinodes in LL$_1$ lead to double peaks in $dI/dV$ curves for each spin polarized LL
such that peak quadruplets act as fingerprints of the two antinodes. We favorably compare our experimental results with numerical calculations \cite{recursive} and with results of an analytic model \cite{Hernangomez}.\\
The experiments were performed in a home-built ultra-high vacuum scanning tunneling microscope (STM) operating at temperature $T=0.4$ K in a magnetic field up to $B=14$ T. The inversion layer is prepared by cleaving p-doped InSb (acceptor density: $ 1\cdot10^{24}/$~m$^3$) and depositing Cs atoms on the surface at $T=40$ K (density: 1.8 \% per InSb(110) unit cell). The dilute density of Cs allows STS of the underlying two-dimensional electron system (2DES). The Cs density is larger than the induced charge density of the 2DES, such that the uncharged Cs atoms effectively screen the minority of positively charged Cs atoms \cite{Morgenstern3}. Thus, the 2DES potential disorder $E_{\rm pot} ({\bf r})$ (${\bf r}$: position) is dominated by the bulk acceptor density \cite{Morgenstern3}.\\
Mapping  $E_{\rm pot} ({\bf r})$  by the spatial dependence of LL$_0$ peaks \cite{Miller} (Fig.\ ~\ref{Fig1}a) reveals a correlation length $\xi=50$ nm and a Gaussian distribution of potential values with sigma-width of 10 meV \cite{Bindel}. Rather isotropic potential valleys are occasionally observed exhibiting diameters up to 30 nm, which is about 4.5 times the magnetic length $l_B = \sqrt{\hbar/eB}=6.8$ nm at $B=14$ T, i.e., such valleys contain five flux quanta. Consequently, the potential is smooth on the length scale of the cyclotron diameter $d_{\rm C} =\sqrt{2n+1}\cdot 2l_B$ for LL$_0$ and LL$_1$. The ratio between confinement energy $\hbar \omega_0$ and LL energy $\hbar \omega_{\rm c}$ of $\omega_0/\omega_{\rm c}\simeq 0.3$ \cite{supplement} implies less than 5 \% mixing between LLs within such valleys \cite{supplement}. An additional mixing of about 10 \% appears due to the relatively strong spin-orbit splitting \cite{supplement} (Rashba parameter $\alpha_{\rm R} \simeq 1$~eV$\rm\AA$ \cite{Becker, Bindel}). \\
Calculating the LDOS for $E_{\rm pot} ({\bf r})$ of Fig.\ ~\ref{Fig1}a by a recursive Green's function algorithm \cite{supplement,recursive}, including the known $\alpha_{\rm R}$, effective mass $m^*\simeq0.03\cdot m_{\rm e}$ ($m_{\rm e}$: bare electron mass), and $g$ factor $g\simeq -21$ \cite{Bindel}, indeed reveals nodal structure for LL$_0$ and LL$_1$ appearing as single stripes and double stripes, respectively (Fig.\ ~\ref{Fig1}b$-$g). The appearance of double and single stripes at the same location and energy with respect to different LL$_n$ centers can be discriminated most easily in areas around smooth potential valleys (arrows in Fig.\ ~\ref{Fig1}c and f). Importantly, the valleys allow to discriminate from arbitrary vicinities of adjacent drift states \cite{Morgenstern2,Hashimoto}, in particular, if the additional overlap of the two spin levels, not regarded in Fig. \ref{Fig1}c$-$g, is taken into account \cite{supplement}.
Notice that albeit the maximum fluctuation of the potential (50 meV) is close to $\hbar \omega_{\rm c}$, different local LLs are always distinct in $dI/dV$ curves (Fig.\ ~\ref{Fig1}h), since $d_{\rm C}\ll\xi$ \cite{supplement}.
An experimental indication of the double stripes is the peak doubling of LL$_1$ as described below.\\
\begin{figure*}[bht]
\includegraphics[width=17.0cm]{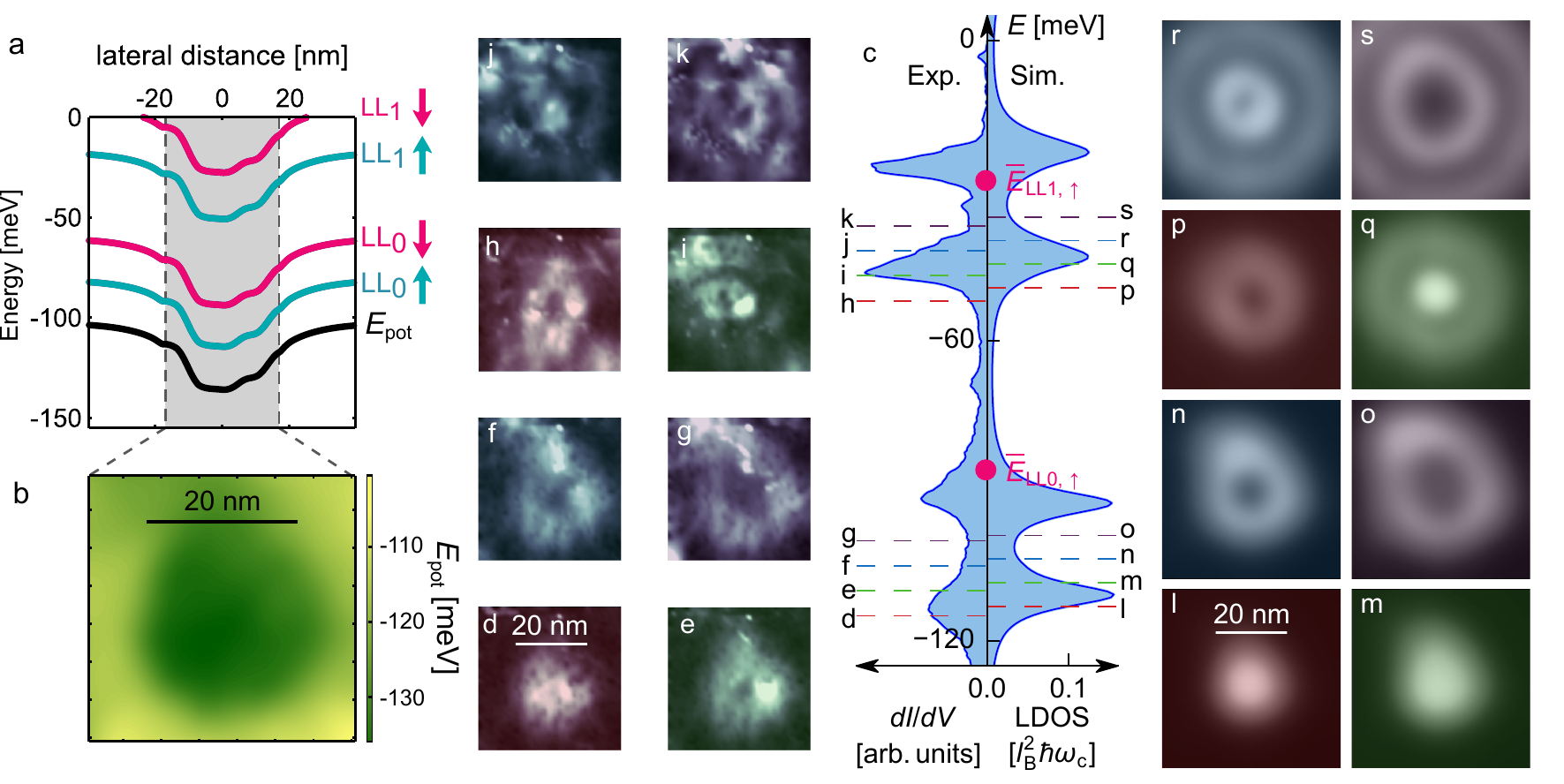}
\caption{(color online).
(a) Cross section through the potential minimum  of b (black) and energy cuts shifted by LL energy and Zeeman energy as indicated by LL$_n$ and spin arrows; (b) $E_{\rm pot} ({\bf r})$ of potential minimum deduced from average LL$_0$ peak energy \cite{Bindel} and subsequently shifted down by $\hbar \omega_{\rm c}/2$; (c) left: $dI/dV$ spectrum recorded in the minimum of the potential valley in b, $B=14$~T, $V_{\text{stab}} = 50~\text{mV}$, $I_{\text{stab}} =150~\text{pA}$, $V_{\text{mod}} = 0.75~\text{mV}_{\text{rms}}$; dashed lines highlight voltages of $dI/dV$ images in d$-$k; right: LDOS$(E)$ in the potential minimum of b calculated by recursive Green's function approach, $m^*=0.028\cdot m_{\rm e}$, $g=-21$, $\alpha_{\rm R} =0.5$~eV$\rm\AA$, $B=14$~T; the LDOS peaks are broadened by a Lorentzian function representing a life time broadening of 0.05 $\hbar \omega_c$ in order to mimic the smallest peak width found within the experiment; dashed lines highlight the energies of the LDOS images in l$-$s; pink dots labeled $\overline{E}_{{\rm LL}n,  \uparrow}$ mark the average LL energy obtained in the area of Fig.\ ~\ref{Fig1}a; (d)-(k) $dI/dV$ images recorded at the voltages marked in c; same parameters as c; (l)$-$(s) calculated LDOS images at the energies marked in c; same parameters as c; image sizes of d$-$k and image sizes of l$-$s are identical.
\label{Fig2}}
\end{figure*}
Concentrating on the widest potential valley (Fig.\ ~\ref{Fig2}) found during several potential scans and using a tip with negligible tip-induced band bending \cite{supplement}, we experimentally observe the development of single and double stripes in real space.
Figure~\ref{Fig2}a and b display the experimentally obtained $E_{\rm pot} ({\bf r})$. The left side of Fig.\ ~\ref{Fig2}c shows a $dI/dV$ curve mapped in the minimum of this potential valley. Four distinct peaks are visible. They correspond to the two spin levels of LL$_0$ and LL$_1$ as verified by comparing with the known $m^*\simeq 0.03\cdot m_{\rm e}$ and $g\simeq -21$ \cite{Bindel,Becker} and seen rather directly by line scans of $dI/dV(V)$ \cite{supplement}. The pink points mark the average LL energy $\overline{E}_{{\rm LL}n, \uparrow}$ resulting from averaging over larger areas as shown in Fig.\ ~\ref{Fig1}a. It is aligned with the average peak voltage of LL$_{0,\uparrow}$ in experimental $dI/dV$ curves from areas of (200 nm)$^2$ \cite{supplement}. Obviously, the peaks within the center of the potential valley are downshifted by about 25 meV with respect to $\overline{E}_{{\rm LL}n, \uparrow}$ indicating confinement (also apparent within line scans of $dI/dV(V)$ \cite{supplement}). The $dI/dV$ images at the voltages marked by dashed lines are shown in Fig.\ ~\ref{Fig2}d$-$k (more detailed energy sequence in Fig. 3 of supplement \cite{supplement}). They are all acquired at energies below $\overline{E}_{{\rm LL}n, \uparrow}$, thus being confined by the potential valley \cite{supplement}. For LL$_{0,\uparrow}$ (Fig.\ ~\ref{Fig2}d-g), one observes that a central disk develops into a ring structure increasing in diameter with increasing voltage. The full width at half maximum of the ring is about 8 nm, i.e. close to $l_B$. This represents the expectation for drift states of LL$_0$, which map equipotential lines at a resolution of $l_B$ \cite{Ando,Morgenstern2,Hashimoto}. Within LL$_1$, at similar energies with respect to $\overline{E}_{{\rm LL}1, \uparrow}$, one observes the development from a small ring structure into a double ring structure growing in size with increasing energy (Fig.\ ~\ref{Fig2}h$-$k). The average distance between the inner and the outer ring in Fig.\ ~\ref{Fig2}k amounts to $11.6\pm 0.3$ nm being slightly smaller than the expected distance of parallel lines within the first LL wave function: $\Delta_{{\rm LL1}} \simeq 2\cdot l_{\rm B}=13.6$ nm \cite{Landau}. Moreover, the smallest structure found in the first LL is a ring, which gets not closed at lower energy, as expected \cite{Landau} (see also detailed sequence in Fig. 3 of supplement).\\
We compare the
experimental results with the same straightforward calculation of the LDOS as used for Fig.\ ~\ref{Fig1}b$-$g, but we adapt $\alpha_{\rm R}$ and $m^*$, which are known to spatially fluctuate \cite{Bindel}, in order to reproduce the $dI/dV$ curve. The resulting LDOS($E$) in the same potential minimum is shown in Fig.\ ~\ref{Fig2}c on the right and corresponding LDOS plots at the marked energies in Fig.\ ~\ref{Fig2}l$-$s.
The general symmetries of the LDOS(${\bf r}$) are well reproduced. However, the calculated structures are larger and grow faster in size with increasing energy. Moreover, the distance between the rings in Fig.\ ~\ref{Fig2}r and s is now rather exactly $2\cdot l_B$.
We believe that the smaller size in the experiments is firstly caused by the fact that we cannot probe details of $E_{\rm pot} ({\bf r})$ at length scales below $l_B$, such that the potential is probably deeper than displayed
in Fig.\ ~\ref{Fig2}b. This leads to additional compression of the wave function not considered in Fig.\ ~\ref{Fig2}l$-$s, namely a more effective rescaling of the magnetic length by $(1+4(\omega_0/\omega_{\rm c})^2)^{-1/4} \simeq 0.9$ \cite{supplement}.
Moreover, $m^*(E)$ of InSb increases with $E$ \cite{Merkt} and since $\omega_0/\omega_c \propto \sqrt{m^*}$, the influence of the potential curvature on the wave functions increases with $E$. This leads to an additional effective compression of the wave functions at high energy not considered in the calculation, which assumes $m^*(E)=$~const..\\
\begin{figure}[tb]
\includegraphics[width=8.5cm]{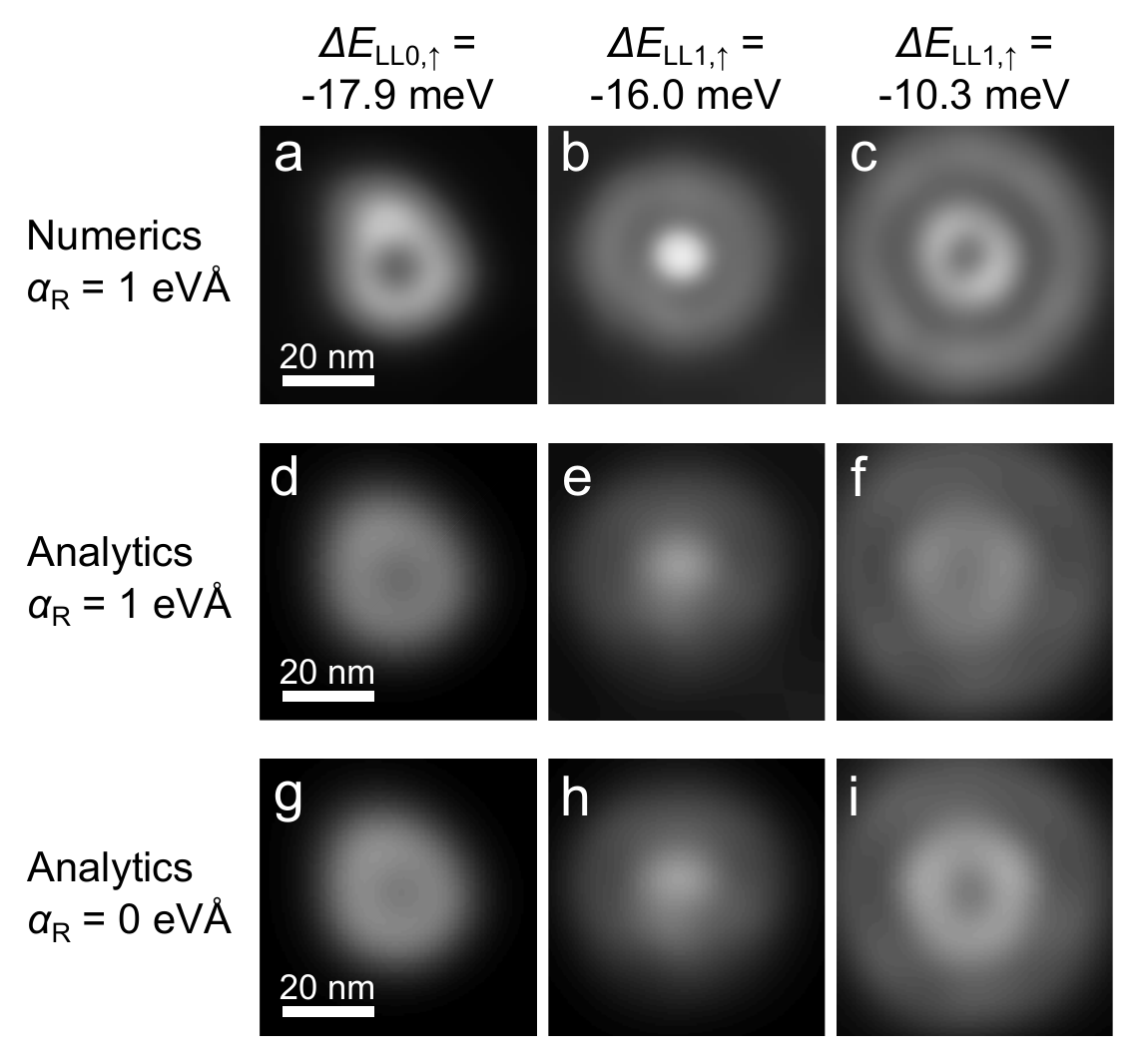}
\caption{(color online).
(a)$-$(c)  LDOS images at the energies marked on top as calculated for $E_{\rm pot} ({\bf r})$ of Fig.\ ~\ref{Fig2}b by recursive Green's function approach; parameters as in Fig.\ ~\ref{Fig1}b$-$g;
(d)$-$(f) LDOS images using the same potential and calculated by the analytic guiding center approach including $\alpha_{\rm R}$ \cite{Hernangomez}; same parameters as in a$-$c;
(g)$-$(i) same as d$-$f but neglecting $\alpha_{\rm R}$.
\label{Fig3}}
\end{figure}
In order to highlight the generic properties of the observed nodal structure, Fig.\ ~\ref{Fig3} compares different calculation schemes of the LDOS$({\bf r})$ for selected $E$. The first line originates from the calculations also presented in Fig.\ ~\ref{Fig2}l$-$s. The second line uses the analytic guiding center approach including the Rashba type spin-orbit interaction \cite{Hernangomez}, while the third line is the most generic description using the guiding center approach without Rashba spin-orbit coupling, i.e. \cite{Ando,Hernangomez}:
%
\begin{align}
{\rm LDOS}(E,{\bf r}) = \sum_{n=0}^\infty \sum_{s=-1/2}^{1/2} \frac{(-1)^{n+1}}{4\pi^2 l_B^4}\nonumber\\
\int n_{\rm F}(E-\epsilon_{n,s}-V_n({\bf R})) \nonumber\\
 L_n\left[ \frac{2(\bf{r}-\bf{R})^2}{l_B^2} \right] \cdot e^{-\frac{({\bf r}-{\bf R})^2}{l_B^2}}\hspace{0.5mm} d^2\bf{R}
\end{align}
%
with $n_F(E)$ being the derivative of the Fermi function with respect to energy, $L_n$ being the $n$th Laguerre polynomial, $\epsilon_{n,s}= (n+1/2)\hbar \omega_c + s g \mu_B B$ ($s$: spin quantum number, $\mu_B$: Bohr's magneton), and:
\begin{equation}
V_n({\bf R}) = \frac{(-1)^n}{\pi l_B^2} \int E_{\rm pot}({\bf r})\cdot L_n\left[ \frac{2({\bf r}-{\bf R})^2}{l_B^2} \right] \cdot e^{-\frac{({\bf r}-{\bf R})^2}{l_B^2}} \hspace{0.5mm}d^2 {\bf r}
\end{equation}
The  generic features, in particular, the double stripe structure (Fig.\ ~\ref{Fig3}c, f, i) are barely changed for the different calculations. They appear most clearly within the numerics allowing additional wave function mixing and, thus, a rescaling of the effective $l_B$ as mentioned above. Thus, the potential valley itself helps to fit the generic double stripe structure into the valley size. The result including $\alpha_{\rm R}$ is slightly more blurred than without $\alpha_{\rm R}$ reflecting the well known mixing of adjacent LLs by the spin-orbit interaction \cite{Hernangomez}.\\
\begin{figure*}[tb]
\includegraphics[width=17cm]{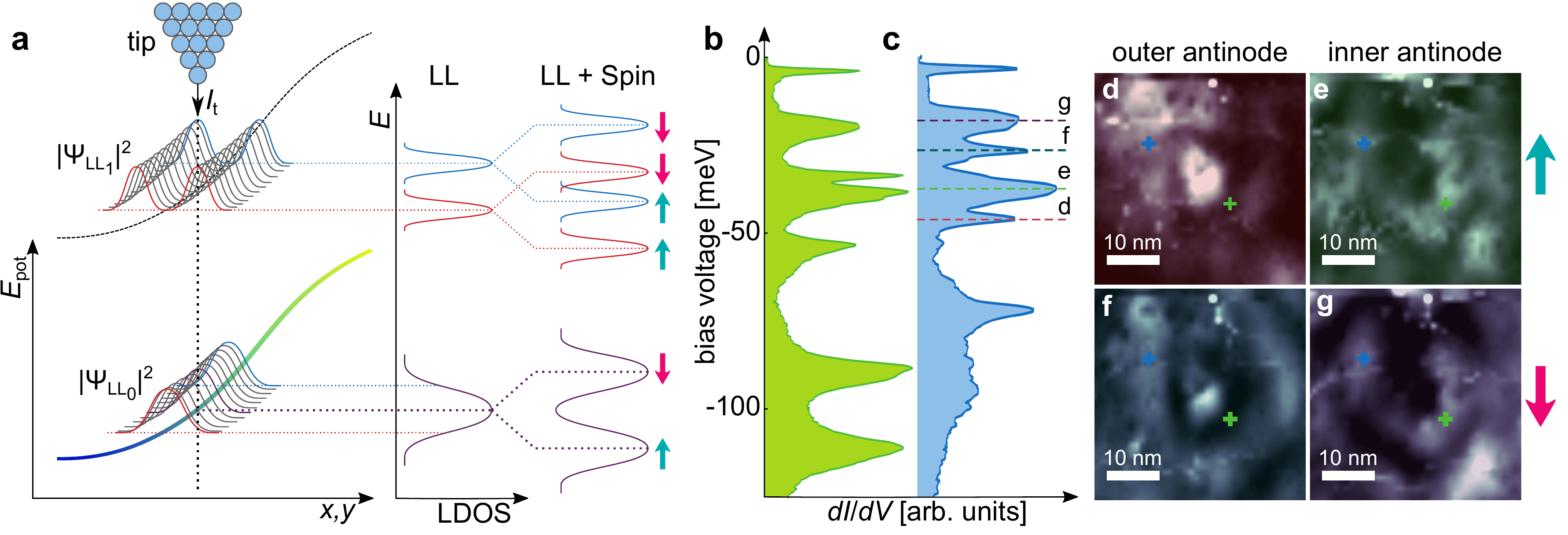}
\caption{(color online).
(a) Sketch of the squared wave functions $|\Psi(x,y)|^2$ of LL$_0$ and LL$_1$ (grey, blue, red, and violet full lines) within the $E_{\rm pot} ({\bf r})$ (blue-green and black dashed line energetically shifted by $\hbar \omega_{\rm c}$); resulting energies of the LL wave functions are indicated by the equally colored, horizontal, dotted lines cutting the energy axis of the middle plot; the tip (triangle of blue circles) tunnels exclusively into the $|\Psi_{{\rm LL}_n}(x,y)|^2$ at the position marked by a vertical, black dotted line; this results in the spectrum sketched in the middle plot; additional Zeeman splitting leads to doubling of the spectral features as sketched in the right plot; (b), (c) $dI/dV$ spectra recorded at the positions marked by the equally colored crosses in d$-$g, $B=14$~T, $V_{\text{stab}} = 50~\text{mV}$, $I_{\text{stab}} =150~\text{pA}$, $V_{\text{mod}} = 0.75~\text{mV}_{\text{rms}}$; dashed lines mark the voltages of the $dI/dV$ images in d$-$g; (d)$-$(g) $dI/dV$ images at the voltages marked in c; notice the nearly identical pattern and energy splitting of d and f, respectively, e and g.
\label{Fig4}}
\end{figure*}
Finally, we explain the appearance of double peaks for LL$_1$ spin levels within $dI/dV$ curves. Figure~\ref{Fig4}a sketches idealized wave functions $|\Psi|^2$ of LL$_0$ and LL$_1$, having an extension of $d_{\rm C}$ (7 nm, 12 nm), within $E_{\rm pot}({\bf r})$. They exhibit zero and one node perpendicular to the drift path, respectively. Shifting these structures laterally changes their energy approximately by $\langle \Psi|E_{\rm pot}|\Psi\rangle$, i.e. smoothly along $E_{\rm pot} ({\bf r})$. The tip, which probes the 0.2 nm area directly below its apex, can tunnel either in the inner or in the outer anti\-node of a LL$_1$ wave functions. These two wave functions have different energies explaining the double peak structure in $dI/dV$ straightforwardly.\\
Figure~\ref{Fig4}b and c show two experimental spectra obtained at two different positions within a potential valley, both exhibiting four peaks for LL$_1$. Displaying the LDOS images at the four peak energies reveals, firstly, that the patterns are nearly identical for the next-nearest neighbor peaks (Fig.\ ~\ref{Fig4}d and f, respectively, e and g). Thus, the next-nearest neighbor peaks belong to the two spin levels of the same wave function $|\Psi({\bf r})|^2$. The peaks in such a pair, moreover, exhibit nearly identical shapes and are separated by the Zeeman energy of 20 meV. Secondly, the LDOS images demonstrate that the nearest neighbor peaks appear because either the inner ring or the outer ring of the LL wave function is below the tip (blue crosses in Fig.\ ~\ref{Fig4}d$-$g). Accordingly, peaks in the green $dI/dV$ curve (b) at energies of Fig.\ ~\ref{Fig4}e and g belong to the inner ring of the LL wave function (green crosses in Fig.\ ~\ref{Fig4}d,g), while the outer ring is crossing this position at lower energies.

In summary, we demonstrated that the generic nodal structure of LL wave functions can be probed by scanning tunneling spectroscopy of an adsorbate induced 2DES at $B=14$ T, if one concentrates on rather smooth potential valleys. The features of these wave functions are nicely reproduced by a recursive Green's function based calculation, but also by a simplified analytic description within the guiding center approach revealing the generic wave functions directly. Thus the potential valley represents a pinning defect for the LL wave functions very similar to the point defects  which pin Bloch waves at $B=0$ T \cite{Crommie, Morgenstern}. The observation of real-space patterns of LL wave functions can be regarded as an important step towards the observation of more complex, interacting wave functions, e.g., within fractional quantum Hall phases \cite{Tsui} probably being accessible by scanning tunneling microscopy of graphene samples \cite{Dean}.\\
%
%
%
\newpage
{\bf The following material corresponds to the supplement of the publication at Physical Review Letters}
\section{Tip induced band bending}
\begin{figure*}[bth]
\centering
\includegraphics[width=\textwidth]{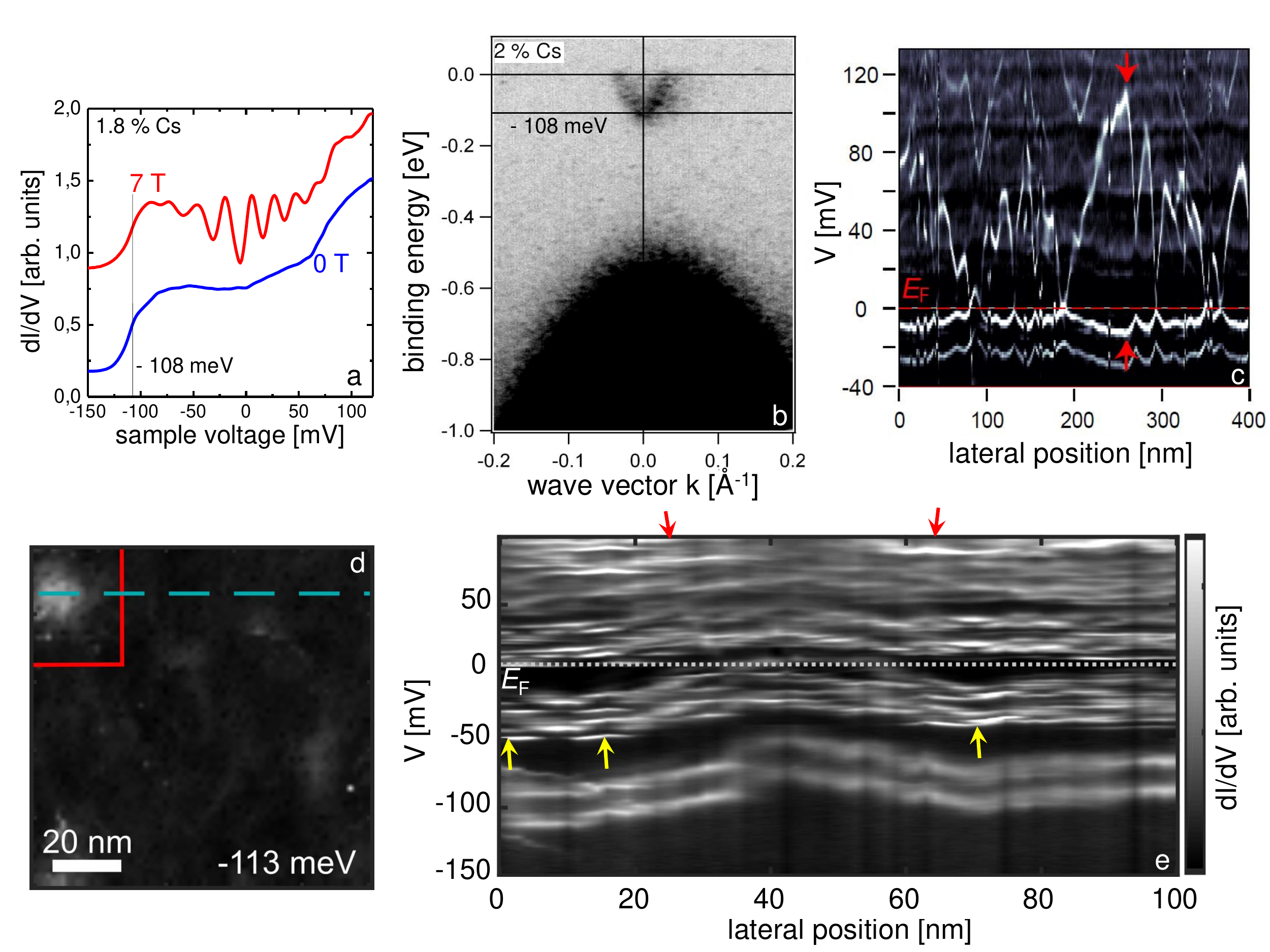}%
\caption{(a) Spatially averaged $dI/dV$ curve of p-InSb(110) covered with 1.8 \% Cs, $B$ as marked, $T=0.4$~K, $V_{\text{stab}} = 300~\text{mV}$, $I_{\text{stab}} =200~\text{pA}$, $V_{\text{mod}} = 3.5~\text{mV}_{\text{rms}}$, $150\times 150$ nm$^2$, $35\times 35$ curves for averaging; the vertical black line labeled $-108$ meV marks the onset of the first subband of the 2DES. \cite{Bindel} (b) ARPES spectrum of the same p-InSb(110) sample covered with 2 \% Cs, photon energy $h\nu=21.2$ eV, $T=80$ K; black vertical line labeled $-108$ meV marks the onset of the first subband of the 2DES. \cite{Morgenstern3} (c) Grey scale plot of $dI/dV$ data along a line of n-InSb(110) covered with 1 \% Cs, $B=8$~T, $T=0.3$~K, $V_{\text{stab}} = 150~\text{mV}$, $I_{\text{stab}} =300~\text{pA}$, $V_{\text{mod}} = 1.0~\text{mV}_{\text{rms}}$. Red arrows mark the appearance of the same state, once if aligned with the Fermi level of the tip (bottom) and once if aligned with the Fermi level of the sample (top) due to tip induced band bending. The stretched mirror appearance of the states can be used to deduce the lever arm $e\Delta V/E_{\rm BB} \simeq 10$.\cite{Hashimoto} (d) Large scale $dI/dV$ image of  p-InSb(110) covered with 1.8 \% Cs, $B=14$~T, $T=0.4$~K, $V=-113$ mV, $V_{\text{stab}} = 50~\text{mV}$, $I_{\text{stab}} =150~\text{pA}$, $V_{\text{mod}} = 0.75~\text{mV}_{\text{rms}}$; red square marks the area probed in Fig. \ref{fig:series} and in Fig. 2-4 of the main text; turquoise dashed line marks the line cut along which $dI/dV$ spectra are presented in e. (e) Grey scale plot of $dI/dV$ along the line marked in d; same parameters as in d; red arrows mark weakly apparent, stretched mirror structures of the states observed close to $E_{\rm F}$; similarly to c, they indicate a lever arm $e\Delta V/E_{\rm BB} \simeq 10$; yellow arrows mark areas where a quadruplet of lines is most clearly observable within LL$_1$; note that a pair of lines of the quadruplet moves exactly parallel indicating the lifted spin degeneracy.
}%
\label{fig:bandbend}%
\end{figure*}
It is well known that tip-induced band bending (TIBB) can influence scanning tunneling microscopy (STM) and scanning tunneling spectroscopy (STS) experiments on III-V semiconductors \cite{Feenstra,Dombrowski}. TIBB is caused by the potential difference between tip and sample which consists of work function differences, typically being up to 400 meV between InSb(110) and a W tip \cite{Dombrowski}, and the applied bias $V$. The most important workaround is to prepare tips with nearly identical work function to the work function of the sample. Therefore, trial and error is applied using cross checks for TIBB. Hereby, the operation at low temperature and in ultrahigh vacuum is decisive, since tips can be kept identical for weeks after successful preparation.\\
We prepared our tips firstly by voltage pulses on W(110) and afterwards by more gentle voltage pulses on InSb(110). The most convincing cross check of residual TIBB is a comparison of the spatially averaged $dI/dV$ spectrum, representing the density of states (DOS), (Fig. \ref{fig:bandbend}a) and the dispersion of the same sample recorded by angular resolved photoelectron spectroscopy (ARPES) (Fig. \ref{fig:bandbend}b).\cite{Morgenstern3} After sufficient pulsing, we achieve negligible differences between the two methods concerning the onset of the first subband of the 2DES, i.e., the difference is less than $\Delta E_{\rm onset}=5$~meV. Hereby, it is important that we operate at a Cs coverage in the saturation range, which starts at 1-2~\%,\cite{Betti,Morgenstern3} such that minor differences in Cs coverage, which cannot be crosschecked by counting the atoms in ARPES, are barely relevant. Moreover, a careful calibration of $E_{\rm F}$ in ARPES and $V=0$ mV in STS is mandatory. We checked, in addition, that the onset energy in STS did barely change with $B$ field (Fig. \ref{fig:bandbend}a, upper curve). \\
A rougher, but faster cross check is given by the absence of confined states of the tip induced quantum dot.\cite{Dombrowski} Such a confined state is easily discernable as a sharp peak below the onset of the first subband of the 2DES, if the TIBB is downwards.\cite{Dombrowski,Morgenstern3} Its energy follows the potential disorder as a function of position.\cite{Morgenstern3} \\
From the more quantitative first cross check, we deduce that the TIBB is negligible at the onset energy of the 2DES. However, it could still be present at different $V$. The strength of the $V$ related TIBB depends on the ratio of the voltage, which drops in vacuum, and the voltage, which drops within the InSb sample. The ratio can be determined experimentally by using the fact that each state crossing the Fermi level of the sample due to TIBB is probed a second time in $dI/dV$ curves.\cite{Hashimoto} Corresponding $dI/dV$ data recorded on n-InSb(110) (doping level $4\cdot 10^{21}/$m$^3$, covered with 1 \% Cs) are shown in Fig. \ref{fig:bandbend}c.\cite{Hashimoto} The red arrows mark a double appearance, once if the tip Fermi level is aligned with the state energy (bottom) and once if the sample Fermi level is aligned with the same state after being shifted by TIBB (top). Stretched mirror lines of the former signals appear due to the latter. They allow to quantify how much tip voltage is required to pull the state up to the Fermi level of the sample by TIBB. For the marked case, it amounts to about 120 mV  (distance between the two white lines marked by the arrows). One straightforwardly deduces the lever arm between the applied additional voltage $\Delta V$ and the shift of the state due to tip induced band bending $E_{\rm BB}$, which is simply the distance of the bottom white line to $V=0$~mV. We deduce $e\Delta V/E_{\rm BB}\simeq 10$.\cite{Hashimoto} For the 1000-fold larger doping used in the actual study, it is rather unlikely that this lever arm is smaller. Instead, one would naively expect the opposite, i.e., a larger lever arm due to the larger screening in the sample. However, the 2DES and the remaining charge in the Cs layer are probably dominating the screening effects, which would lead to a similar value of the lever arms.\\
Figure \ref{fig:bandbend}e shows the same plot as in c for the p-InSb(110) used in this study. The stretched mirror lines are barely visible due to the fact that confined states are also present above the Fermi level of the sample, which is not the case for the degenerately doped n-type sample. However, careful inspection shows remainders of the mirror lines, for states close to $E_{\rm F}$ of the sample, as marked by red arrows. The deduced stretching factor corroborates the result of $e\Delta V/E_{\rm BB}\simeq 10$ also for this sample.\\
\begin{figure}[thb]
\centering
\includegraphics[width=8.2cm]{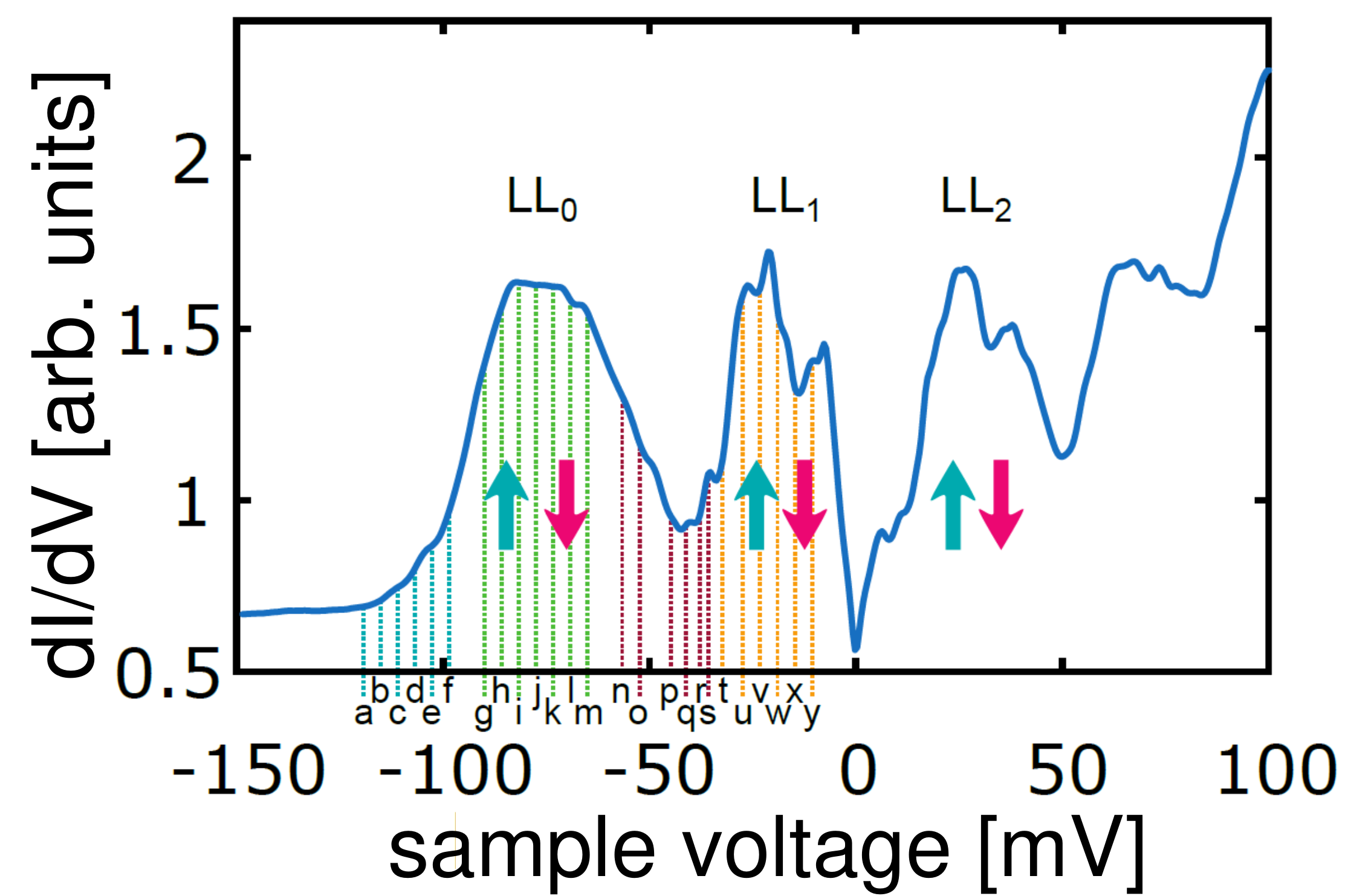}%
\caption{Spatially averaged $dI/dV$ curve of p-InSb(110) covered with 1.8 \% Cs, $B=14$~T, $T=0.4$~K, $V_{\text{stab}} = 300~\text{mV}$, $I_{\text{stab}} =200~\text{pA}$, $V_{\text{mod}} = 3.5~\text{mV}_{\text{rms}}$, $150\times 150$ nm$^2$, $35\times 35$ curves for averaging; different LLs are labelled; arrows mark spin directions; colored vertical lines mark the voltages of $dI/dV$ images in Fig. \ref{fig:series} with identical colors as used for the energy labeling there; note the sharp triangular Coulomb gap at $V=0$ mV, which separates LL$_1$ and LL$_2$ more strongly than LL$_0$ and LL$_1$.}%
\label{fig:spectrum}%
\end{figure}
Consequently, the applied $\Delta V$ in this study, being up to 90 mV above the onset energy of the 2DES (Fig. \ref{fig:series}y), leads to a voltage related TIBB of $\Delta E_V \le 9$ meV. This results in a maximum complete band shift, respectively, a shift of the states by the TIBB of $\Delta E_{\rm BB}=\sqrt{\Delta E_{\rm onset}^2+\Delta E_V^2}\le 10$ meV. Notice, that the potential valley probed in Fig. 2-4 of the main text is about 30 meV in depth such that the remaining $\Delta E_{\rm BB}$ will not change the general confinement property of this valley, but it will slightly stretch the energy scale of the confined states by order 10 \%. The lateral extension of the TIBB might change the shape of the confinement potential slightly, which is probably responsible for the remaining differences between calculated and measured LDOS in Fig. 2 of the main text.
\section{Detailed development of antinodes with energy}
\begin{figure*}[bth]
\centering
\includegraphics[width=0.99\textwidth]{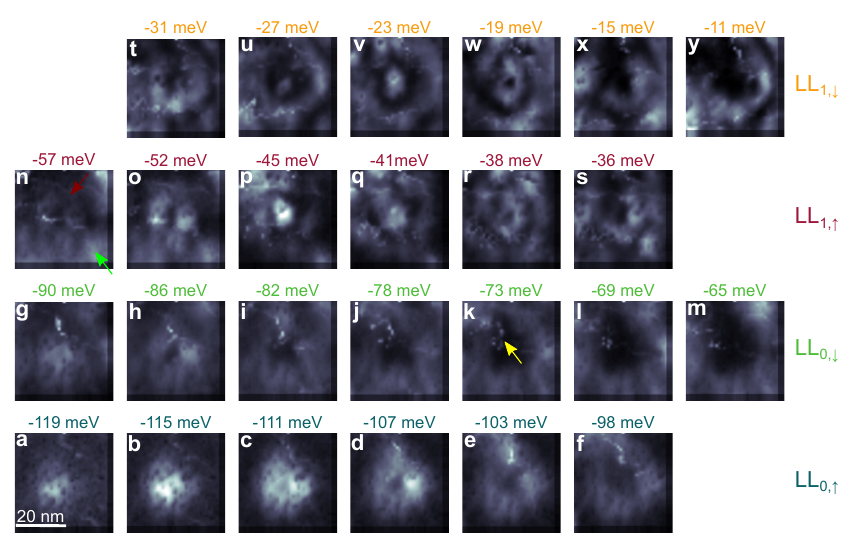}%
\caption{Series of $dI/dV$ images recorded in the area also displayed in Fig. 2$-$4 of the main text; $B=14$~T, $T=0.4$ K, $V_{\text{stab}} = 50~\text{mV}$, $I_{\text{stab}} =150~\text{pA}$, $V_{\text{mod}} = 0.75~\text{mV}_{\text{rms}}$; energies $eV$ (marked in Fig. \ref{fig:spectrum} as colored lines) are labeled on top and constituting LLs are given on the right; green arrow in n marks the remainder of the extended ring from LL$_{0,\downarrow}$, while the outer ring of LL$_{1,\uparrow}$ starts to appear (red arrow); yellow arrow marks one sharp line originating from charging events which partly disturb the LDOS imaging.
}%
\label{fig:series}%
\end{figure*}
Figure \ref{fig:spectrum} shows a spatially averaged $dI/dV$ spectrum obtained at $B=14$ T, which probes the DOS of the complete area displayed in Fig. \ref{fig:bandbend}d and Fig. \ref{fig:longrange}c-f. The spectrum adequately represents the DOS of the 2DES as can be straightforwardly crosschecked by separating the image in smaller pieces and comparing the resulting average $dI/dV$ curve obtained from these smaller areas.  The marked Landau levels (LLs) are clearly separated and spin splitting becomes apparent in LL$_1$ and LL$_2$. Nevertheless, the average $dI/dV$ in between LL$_0$ and LL$_1$ does not drop to the level recorded at voltages below the onset of the 2DES (left area). This indicates a remaining overlap of the DOS of different LLs. However, in real space, spin and Landau levels are always clearly distinct as can be seen in the $dI/dV$  line scan shown in Fig. \ref{fig:bandbend}e. Thus, the fact that the global LLs overlap is caused by the long-range potential disorder with correlation length ($\simeq 50$~nm)\cite{Bindel} much larger than the magnetic length $l_{\rm B}(14$~T$)\simeq 7$~nm. Strong mixing of LLs would only appear, if the potential changes on the scale of the cyclotron radius $r_{\rm c}=\sqrt{2n+1}\cdot l_B$ would be approximately as large as the LL gaps (see section VI for a detailed discussion), which is clearly not the case (see Fig. \ref{fig:bandbend}e).\\
Notice that the $dI/dV$ intensity between LL$_1$ and LL$_2$ reaches the $dI/dV$ level recorded below the onset of the first subband. This is caused by the Coulomb gap, which according to Efros and Shlovskii \cite{Efros} should be linear in energy around $E_{\rm F}$ for localized systems in 2D. The gap has been discussed previously in detail,\cite{Becker1} but the linearity is nicely visible in Fig. \ref{fig:spectrum}, too.\\
The colored lines in Fig. \ref{fig:spectrum} mark the energies of the LDOS images obtained in the potential valley of Fig. 2-4 of the main text as displayed in Fig. \ref{fig:series}. They are colored according to their affiliation to a particular spin polarized LL as deduced from Fig. \ref{fig:series}. Since they belong to the deepest potential valley in that particular area, they are found at the low energy tail of the corresponding LL peak partly penetrating into energy areas which are globally dominated by a lower LL (red lines belong to LL$_{1,\uparrow}$).\\
The LDOS images in Fig. \ref{fig:series} are ordered with respect to their affiliation. One observes for both spin levels of LL$_0$, that a disk like feature representing a Gaussian LDOS develops into a ring increasing in diameter with energy. Eventually, the ring reaches the rim of the potential valley displayed in Fig. 2b of the main text. The structure at the rim is still faintly visible (green arrow in Fig. \ref{fig:series}n), when the outer ring of the LL$_{1,\uparrow}$ wave function (red arrow) appears in the center of the valley. This outer ring does not start as a Gaussian disk at lowest energy, which can be rationalized, e.g., by the fact that it has to be orthogonal to the wave functions of LL$_0$. The ring appears more strongly in Fig. \ref{fig:series}o before starting to increase in diameter with energy and soon being accompanied by a disk in its center evolving into a second, inner ring structure afterwards. It is obvious from this series that the two rings observed in Fig. \ref{fig:series}r$-$s do not contain a remainder of the ring of LL$_{0,\downarrow}$, but are the two antinodes belonging to LL$_{1,\uparrow}$.\\
The procedure of a starting ring increasing in size with energy and soon being accompanied by a disk evolving into a second, inner ring is repeated for LL$_{1,\downarrow}$ (Fig. \ref{fig:series}t$-$y). However, it is difficult to discriminate the outer ring of LL$_{1,\downarrow}$ from the inner ring of LL$_{1,\uparrow}$ at the energies between Fig. \ref{fig:series}s and t (not shown).\\
The $dI/dV$ images in Fig. \ref{fig:series} partly show sharp lines (e.g., yellow arrow in k) in addition to the more smooth LDOS of the LL wave functions. They are attributed to charging of Cs atoms by TIBB as discussed elsewhere.\cite{Teichmann} The Cs atoms itself are visible as small black dots within the probed squared wave functions of the LLs.\cite{Morgenstern3}
\section{Single and double stripes within large scale images}
\begin{figure*}[thb]
\centering
\includegraphics[width=13.5cm]{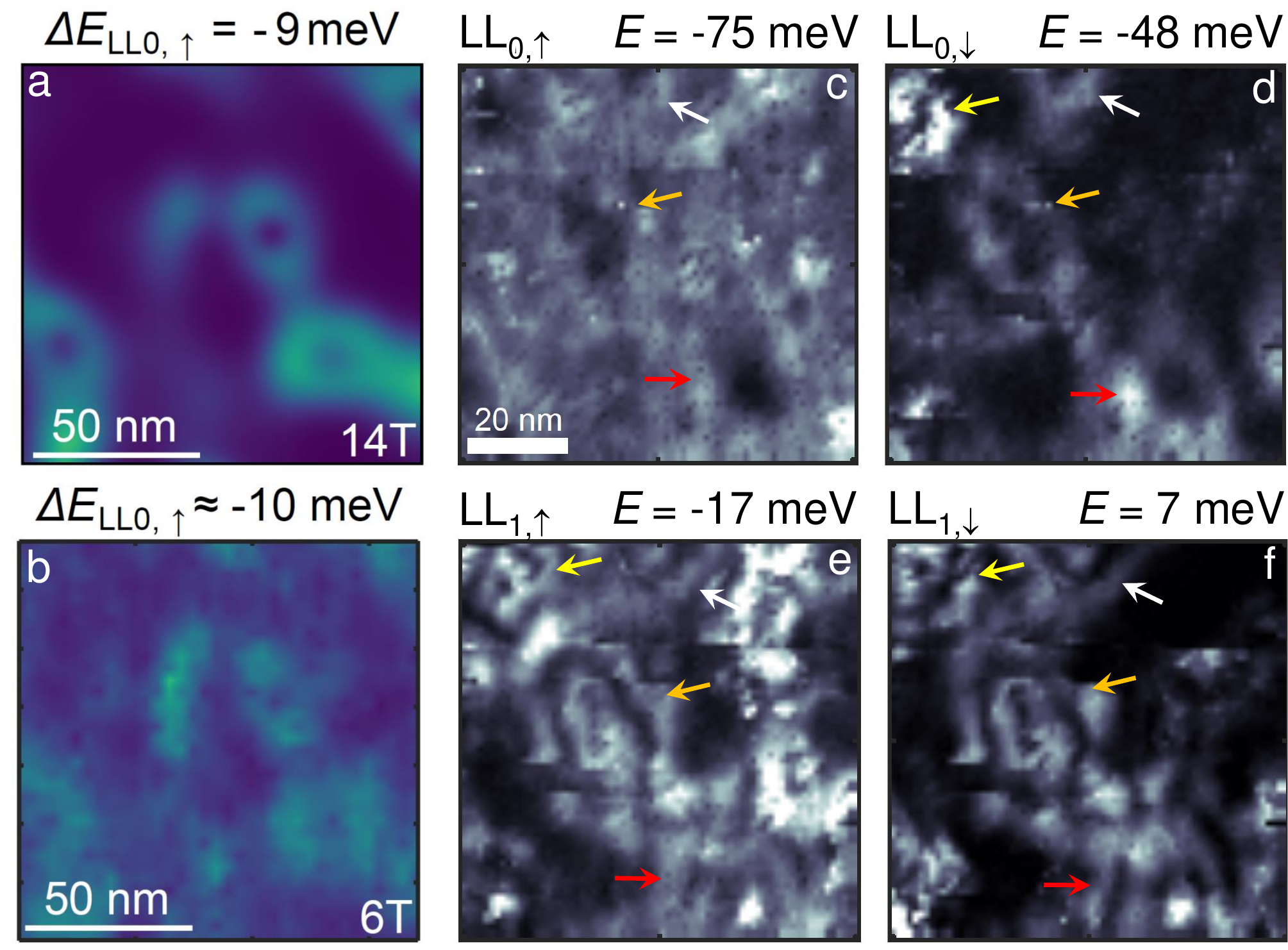}%
\caption{(a) (same as Fig. 1b of the main text) LDOS within LL$_{0,\uparrow}$ as calculated within the recursive Green's function algorithm \cite{recursive} using the potential of Fig. 1a of the main text, $m^*=0.03\cdot m_{\rm e}$ ($m_{\rm e}$: bare electron mass), $g=-21$, $\alpha_{\rm R} =1$~eV$\rm\AA$, $B=14$~T; the energy with respect to the center of the Landau level within the plotted potential area is marked. (b) $dI/dV$ image of the same area, $B=6$~T, $T=0.4$ K, $V_{\text{stab}} = 50~\text{mV}$, $I_{\text{stab}} =100~\text{pA}$, $V_{\text{mod}} = 1.5~\text{mV}_{\text{rms}}$; energy with respect to the average energy of LL$_{0,\uparrow}$ in the displayed area determined from the peak in the averaged $dI/dV$ curve (similar to Fig. \ref{fig:spectrum}) is marked on top; note the similar LDOS structures in a and b and the larger width of the structures in b due to the lower $B$, hence, larger $l_B$. (c)$-$(f) Large scale $dI/dV$ images of another area at energies $E=eV$ as marked with the corresponding LL on top, $B=14$~T, $T=0.4$ K, $V_{\text{stab}} = 50~\text{mV}$, $I_{\text{stab}} =150~\text{pA}$, $V_{\text{mod}} = 0.75~\text{mV}_{\text{rms}}$; identically colored arrows point to the same area in different images, where the development from a single stripe (d, partly c) to double stripes (e,f) of the drift states is most obvious.}
\label{fig:longrange}%
\end{figure*}
The calculated LDOS of Fig. 1b$-$g of the main text implies that double stripes for LL$_1$ appear at areas of single stripes of LL$_0$ also on more flat areas of the potential. This is most obvious for Fig. 1d and g, but also discernable by comparing Fig. 1c and f.
These more extended structures are more difficult to discriminate within the experiment since, unlike the
spin-resolved numerical LDOS data, the experimental dI/dV data show spatially overlapping contributions from both spin levels. This is why we concentrate on the deep potential valley in the main text, which allows more quantitative comparison.\\
However, more extended structures evolving from single lines to double lines can also be found in experiment at the upper rim of the DOS of a particular LL as demonstrated in Fig. \ref{fig:longrange}c$-$f. A number of structures are marked by arrows which nicely show the doubling of lines in LL$_1$ (e, f) with respect to LL$_0$ (c, d), in particular, if one considers the two upper spin levels in d and f. In c and e, the overlap of the DOS of the two spin levels becomes apparent by the structures in the upper right area. These structures do not exhibit any doubling between c and e and are not observable in d and f. These LDOS structures, belonging to LL$_{1,\downarrow}$ in e, correspond to the single outer ring seen also in Fig. \ref{fig:series}t for the deep potential valley, i.e., they represent the low-energy single lines of LL$_{1,\downarrow}$ within shallower potential valleys, which will develop into a double line structure at larger energies.\\
 Unfortunately, the $dI/dV(x,y,V)$ data set leading to Fig. \ref{fig:longrange}c$-$f contains some instabilities, possibly due to minimal tip changes, which make it impossible to deduce a reliable potential map such that a direct comparison with the calculated LDOS is not at hand. Instead, we show such a comparison in Fig. \ref{fig:longrange}a$-$b albeit only for LL$_0$. The similarity of the LDOS structures in experiment and calculation is apparent, where the measured structures obtained at $B=6$ T are slightly broader than the calculated ones obtained at $B=14$ T in line with the different $l_B\propto B^{-0.5}$ determining its width.\cite{Morgenstern2}
\vspace{10cm}
\section{Quadruplets of peaks in LL$_1$}

\begin{figure}[thb]
\centering
\includegraphics[width=8.5cm]{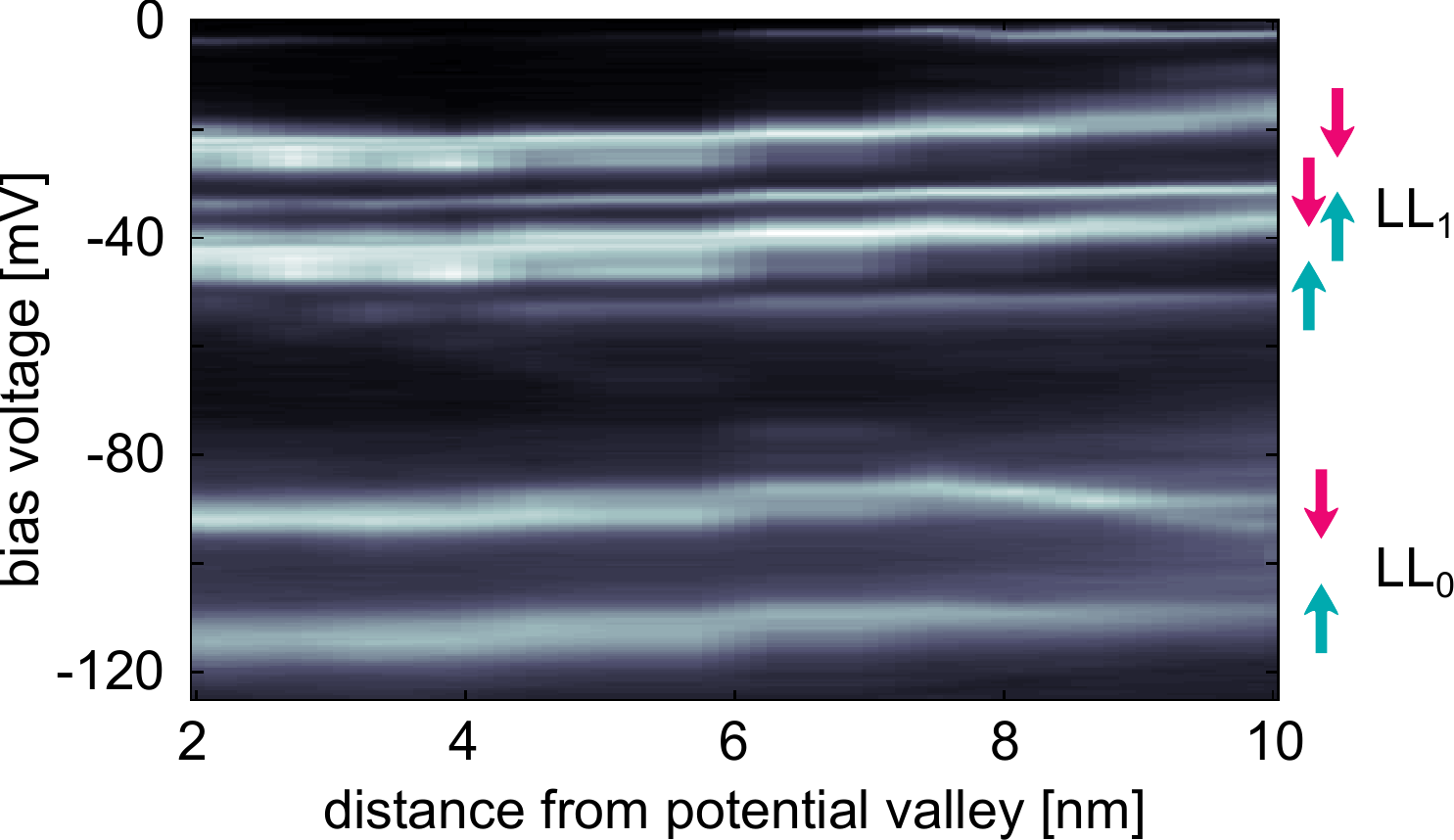}%
\caption{Grey scale plot of $dI/dV$ intensity along a line of the potential shown in Fig. 2a of the main text. Landau levels LL$_n$ and spin levels (arrows) are marked, $B=14$~T, $V_{\text{stab}} = 50~\text{mV}$, $I_{\text{stab}} =150~\text{pA}$, $V_{\text{mod}} = 0.75~\text{mV}_{\text{rms}}$.}%
\label{fig:quadruplet}%
\end{figure}
Figure \ref{fig:quadruplet} shows the $dI/dV$ intensity along a line within the intermediate area of the potential of Fig. 2b of the main text. It is obvious that LL$_1$ consists of four states which pairwise behave very similar along the line. Each next nearest neighbor pair belongs to the different spin levels of one anti\-node corroborating our assignments from the main text.
Fig. \ref{fig:bandbend}e shows further examples of such quadruplets of $dI/dV$ lines as marked, e.g., by the yellow arrows. They are most clearly seen in relatively flat potential areas away from potential minima, where the model discussed in Fig. 4a of the main text applies most favorably, i.e. the potential curvature on the scale of $l_B$ is negligible.\cite{Champel,Hernangomez}

\vspace{7cm}
\section{Mixture of spin-channels due to Rashba spin-orbit coupling}
\begin{figure}[t]
\centering
\includegraphics{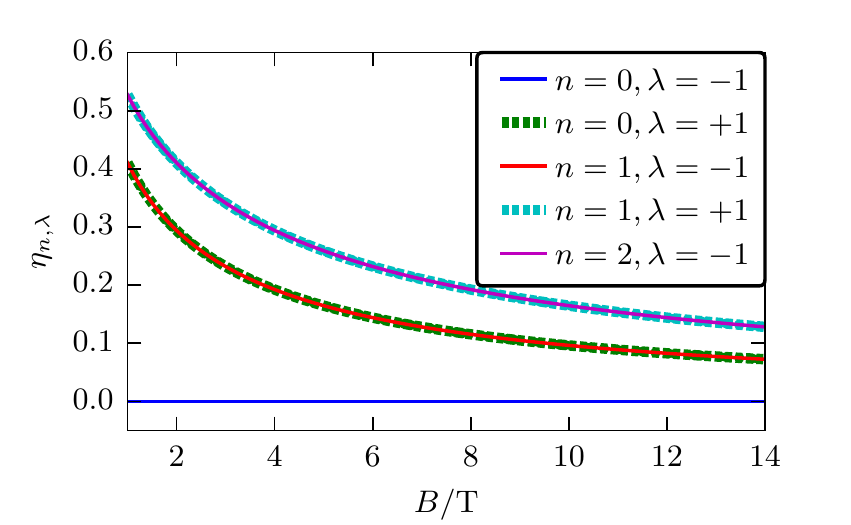}%
\caption{Plot of the mixing parameter $\eta_{n,\lambda}$ from
	Eq.~\eqref{eq:mixing_parameter}, which describes the relative weight
	of the spin-polarization $-\lambda$ in the state $|n,\lambda\rangle$,
	as a function of magnetic field $B$. The parameters are $m=0.03\,m_e$, $\alpha_R = 1\, \mathrm{eV \AA}$, $g = -21$.
}%
\label{fig:mixing}%
\end{figure}

The Rashba Hamiltonian
including the Zeeman interaction and a disorder potential $V(\vec r)$ reads
\begin{align}\label{eq:ham_rashba_bare}
H = \frac{\vec{\pi}^2}{2m} + \frac{\alpha_R}{\hbar}(\pi_x \sigma_y - \pi_y \sigma_x)
+ \frac{1}{2} g \mu_B B \sigma_z + V(\vec r),
\end{align}
where the first term is the kinetic energy corresponding to the physical
momentum $\vec \pi = \vec p + e \vec A(\vec r)$ with $e > 0$ the electron
charge and the vector potential $\vec A(\vec r)$, the second term is the Rashba
interaction with the Pauli matrices $\sigma_{x,y,z}$ and Rashba parameter
$\alpha_R$, and the third term is the Zeeman interaction with the electron
$g$-factor $g$, the Bohr magneton $\mu_B$ and the magnetic field strength $B$.
In absence of disorder with $V(\vec r) = 0$, the eigenstates of the
Hamiltonian~\eqref{eq:ham_rashba_bare} can be written in the
form~\cite{Bychkov, Winkler}
\begin{align}
|n, \lambda\rangle
&= c^{(1)}_{n,\lambda} |n\rangle \otimes |\sigma_\lambda\rangle + c^{(2)}_{n,\lambda} |n +
\lambda \rangle \otimes |\sigma_{-\lambda}\rangle,
\end{align}
where the states $|\sigma_\lambda\rangle$ are eigenstates to spin polarization
$\lambda$ such that $\sigma_z |\sigma_\lambda\rangle = \lambda
|\sigma_\lambda\rangle$ and the states $|n\rangle$ are states in Landau level
$n$.  The coefficients $c^{(1/2)}_{n,\lambda}$ are given by
\begin{align}
c^{(1)}_{n,\lambda} &= \cos \theta_{n+(1+\lambda)/2},
&
c^{(2)}_{n,\lambda} &= \lambda \sin \theta_{n+(1+\lambda)/2},
\end{align}
and depend on the angles
\begin{align}\label{eq:rashba_angle}
\theta_n &= \arctan \Biggl[ \frac{\sqrt{n} S}{1-Z + \sqrt{(1-Z)^2 +
		nS^2}}\Biggr]
\end{align}
which parametrize the strength of the contributions to different spin
components. Here, $Z = g \mu_B B/\hbar \omega_c$ and $S = \sqrt{2} 2
\alpha_R/\hbar \omega_c l_B$ measure the strength of the Zeeman interaction and
the Rashba interaction in terms of the cyclotron frequency $\omega_c = eB/m$
and the magnetic length $l_B = \sqrt{\hbar/eB}$. The eigenenergies are obtained
as
\begin{align}\label{eq:rashba_eigenenergies}
E_{n,\lambda} = \hbar \omega_c \Bigl[ n + \tfrac{1+\lambda}{2}
- \tfrac{\lambda}{2} \sqrt{(1-Z)^2 + (n+\tfrac{1+\lambda}{2}) S^2} \Bigr].
\end{align}

The above choice of labeling of the states has the advantage that in the limit $S
\rightarrow 0$, the state $|n,\lambda\rangle$ smoothly goes over to a state in
Landau level $n$ with spin polarization $\lambda$ . Indeed, from
Eq.~\eqref{eq:rashba_angle} it is obvious that for $S \rightarrow 0$, we find
$\theta_n \rightarrow 0$ and correspondingly, $c_{n,\lambda}^{(2)} \rightarrow
0$. It is important to note that $S \rightarrow 0$ as $B \rightarrow \infty$,
such that the effective Rashba spin-orbit coupling gets weaker as the magnetic field
increases. At large magnetic fields, we expect that we can approximately regard
the states $|n,\lambda\rangle$ as states in Landau level $n$ with spin
polarization $\lambda$. To characterize the remaining contribution to the wave
function $|n,\lambda\rangle$ in the spin channel $-\lambda$ it is convenient to
define the mixing parameter $\eta_{n,\lambda} =
|c_{n,\lambda}^{(2)}|^2/|c_{n,\lambda}^{(1)}|^2$ as the ratio of the weights of
both spin polarizations in the square of the wave-function $|\psi(\vec r)|^2$.
This means that the contribution to the local density of states by the spin
polarization $-\lambda$ in the state $|n,\lambda\rangle$ is smaller by a factor
of $\eta_{n,\lambda}$ compared to the contribution of spin-polarization
$\lambda$. The mixing parameter $\eta_{n,\lambda}$ can be written explicitly as
\begin{align}\label{eq:mixing_parameter}
\eta_{n,\lambda}
= \frac{[n+(1+\lambda)/2] S^2}{\Bigl(1-Z + \sqrt{(1-Z)^2 +
		\bigl[n+\tfrac{1+\lambda}{2}\bigr] S^2}\Bigr)^2},
\end{align}
which shows that every state $n,\lambda=+1$ has a contribution in the spin-down
component which is of the same intensity as the contribution in the spin-up
component of the state $n+1$, $\lambda=-1$. The mixing parameter is shown in
Fig.~\ref{fig:mixing} as a function of magnetic field for the parameter values
$m=0.03\,m_e$, $\alpha_R = 1\, \mathrm{eV \AA}$, $g = -21$. In particular, for
$B = 14\,\mathrm{T}$, we obtain the mixing parameters $\eta_{1,\lambda=-1}=\eta_{0,\lambda=1}
\approx 0.07$, $\eta_{1,\lambda=1} \approx 0.13$, showing that the mixture
of Landau levels LL$_0$ and LL$_1$ due to Rashba spin-orbit coupling indeed
remains small for our parameters.

\section{Landau-level mixing due to the potential valley}
Having seen that spin-orbit coupling has only a negligible effect, we now want
to assess the importance of Landau-level mixing due to the potential. To that
end, we consider the spinless Fock-Darwin problem
\begin{align}
H = \frac{\vec \pi^2}{2 m} + \frac{1}{2} m \omega_0^2 \vec r^2
\end{align}
of electrons in the $xy$-plane subject to a magnetic field $\vec B = B \vec
e_z$ in the $z$-direction and a quadratic confinement potential with
characteristic frequency $\omega_0$. The Fock-Darwin Hamiltonian possesses
exact solutions\cite{Fock}
\begin{align}
	\psi_{n,l}(\vec r) = \sqrt{\frac{n!}{(n+|l|)!}} \biggl(
	\frac{r}{\sqrt{2} L_B} \biggr)^{|l|}\nonumber \\
	\times L_n^{|l|}\biggl(\frac{r^2}{2 L_B^2}\biggr)
e^{-r^2/4 L_B^2} \frac{e^{-i l \phi}}{\sqrt{2\pi L_B^2}},
\end{align}
where $n \in \mathbb{N}_0$ and $l \in \mathbb{Z}$ are quantum numbers and we
use the radius $r = |\vec r|$ and the angle $\phi$ in polar coordinates. Since
the electric confiment potential adds to the confinement produced by the
magnetic field, the magnetic length $l_B = \sqrt{\hbar/m \omega_c}$ is
renormalized to $L_B = \sqrt{\hbar/m \Omega}$ with the frequency $\Omega =
\sqrt{\omega_c^2 + 4 \omega_0^2}$.

We observe that finite $\omega_0$ decreases the length scale on which the
wave function varies from $l_B$ to $L_B$, but leaves the overall form of the
wave function invariant. The reason for this is that the quadratic potential
leaves the harmonic structure of the Hamiltonian without confinement potential
intact. To see this more explicitly, it is instructive to introduce new quantum
numbers $N_\chi \in \mathbb{N}_0$, $M_\chi \in \mathbb{N}_0$, which are related
to $n$, $l$ as $l = N_\chi-M_\chi$, $n = (M_\chi + N_\chi -
|N_\chi-M_\chi|)/2$. The reason for the subscript $\chi$ will become apparent
below. In terms of the new quantum numbers, the Fock-Darwin spectrum assumes
the form
\begin{align}
E = \hbar \Omega
\bigl[ n + (|l|+1)/2\bigr] + \frac{\hbar \omega_c}{2} l \\
= \hbar \frac{\Omega + \omega_c}{2} (N_\chi+1/2)\label{eq:fock_darwin_spectrum}
+ \hbar \frac{\Omega - \omega_c}{2} (M_\chi+1/2).
\end{align}
This makes the harmonic structure of the renormalized Landau levels (associated
with $N_\chi$) and the renormalized guiding center levels (associated with
$M_\chi$) explicit.

The average radius of the wave-function scales as
\begin{align}\label{eq:radius_avg}
	\sqrt{\langle \vec r^2 \rangle} = \sqrt{2 L_B^2}(N_\chi + M_\chi + 1),
\end{align}
showing that the wave functions become more extended as $N_\chi$ and $M_\chi$
increase.  Therefore, the mismatch between a wave function varying on the
larger scale $l_B$ and the exact wave function varying on the smaller scale
$L_B$ becomes larger and Landau level mixing increases with $N_\chi$ and
$M_\chi$.  However, since Landau-level mixing does not affect the geometric
structure of the wave function, we expect that even a description neglecting
Landau-level mixing will capture the overall wave-function structure.

To quantify the strength of Landau-level mixing, it is
convenient to introduce the cyclotron coordinates $\eta_y = -\pi_x/m \omega_c$,
$\eta_x = \pi_y/m \omega_c$ and the guiding center coordinates $X = x -
\eta_x$, $Y = y-\eta_y$.
Since the coordinates are canonically conjugate, $[X,Y] = i l_B^2$ and
$[\eta_x, \eta_y] = -i l_B^2$, one can introduce ladder operators
\begin{align}
a &= (\eta_x - i \eta_y)/\sqrt{2}l_B
&
b &= (X+iY)/\sqrt{2}l_B,
\end{align}
which have the usual nonvanishing commutation relations $[a,a^\dag] =
[b,b^\dag] = 1$. Here, $a$ is related to the Landau level index and $b$ is
related to the guiding center position. Expressing the Hamiltonian in terms of
ladder operators leads to
\begin{align}
H = \hbar \omega_c (a^\dag a + \tfrac 1 2) + \frac{\hbar \omega_0^2}{\omega_c} (
a^\dag a + b^\dag b + a^\dag b^\dag + a b + 1).
\end{align}
The presence of terms quadratic in the creation and annihilation operators
suggests that diagonalizing the system requires a Bogoliubov transformation.
Indeed, one can verify that the Hamiltonian is diagonalized by a unitary
transformation
\begin{align}\label{eq:fock_darwin_unitary}
U = \exp\bigl[\chi( a b - a^\dag b^\dag)\bigr]
\end{align}
with
\begin{align}
	\chi = \frac{1}{4} \log\biggl(1+ \frac{4 \omega_0^2}{\omega_c^2}\biggr).
\end{align}
The diagonalized Hamiltonian $H_\chi = U H U^\dag$ reads
\begin{align}\label{eq:fock_darwin_diagonalized}
H_\chi &= \hbar \frac{\Omega}{2} \bigl[ a^\dag_\chi
a_\chi +
b^\dag_\chi b_\chi + 1\bigr]
+ \frac{\hbar \omega_c}{2}\bigl[a^\dag_\chi a_\chi -
b^\dag_\chi b_\chi\bigr]
\end{align}
in terms of the transformed operators $a_\chi = U a U^\dag$, $b_\chi = U a
U^\dag$, which are given by
\begin{align}
a_\chi = a \cosh \chi + b^\dag \sinh \chi \\
b_\chi = b \cosh \chi + a^\dag \sinh \chi.
\end{align}
Denoting by $N_\chi$, $M_\chi$ the number of quanta of $a_\chi$ and $b_\chi$,
respectively, we recover the form of the Fock-Darwin spectrum given in
Eq.~\eqref{eq:fock_darwin_spectrum}.

The ground state $|0_\chi,0_\chi\rangle$ of the
Hamiltonian~\eqref{eq:fock_darwin_diagonalized} is defined by the condition
$a_\chi |0_\chi,0_\chi\rangle = b_\chi|0_\chi,0_\chi\rangle = 0$. It is related to the vacuum
$|0,0\rangle$ of $a$ and $b$ defined by $a |0,0\rangle = b |0,0\rangle = 0$
through the unitary transformation~\eqref{eq:fock_darwin_unitary}. The unitary
transformation~\eqref{eq:fock_darwin_unitary} performs a two-mode
squeezing of the vacuum $|0,0\rangle$. Using standard results about two-mode
squeezing~\cite{Vogel}, it is straightforward to calculate the overlap of
states $|N_\chi, M_\chi\rangle$ with the states $|N,M\rangle$ associated with
the bare Landau level $N$ and guiding center state $M$ that is not renormalized
by $\omega_0$. Using this, we compute the spectral weight of the state
$|N_\chi=N, M_\chi\rangle$ in Landau level $N$. We obtain
\begin{align}
W_N &= \sum_{M} |\langle N,M |  N_\chi=N, M_\chi \rangle|^2 \nonumber \\
&= \label{eq:form1}
\frac{F(-N, -M_\chi, 1, - \sinh^2\chi)^2}{\cosh(\chi)^{2N+2 M_\chi+2}},
\end{align}
where $F(a,b,c,z)$ is the hypergeometric function.

\begin{figure}[tb]
\centering
\includegraphics{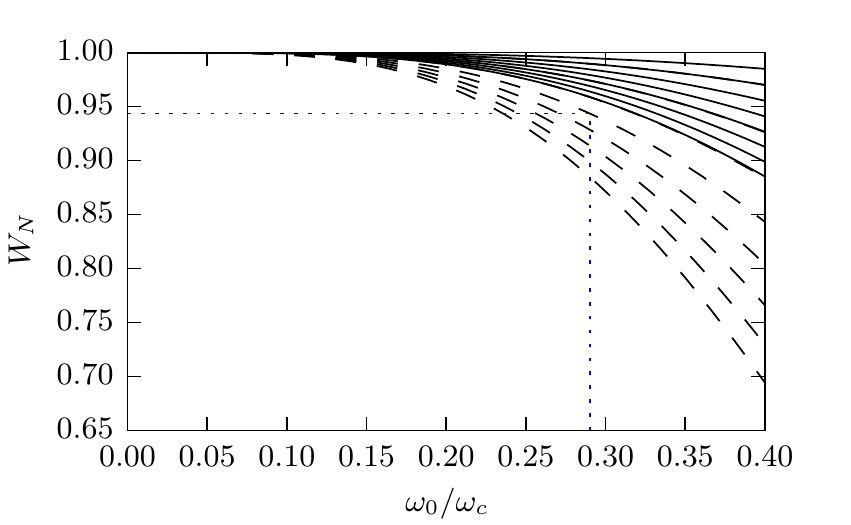}%
\caption{Spectral weight of the states $|N_\chi=N,M_\chi\rangle$ in the bare
	Landau levels $N=0$ (solid lines) and $N=1$ (dashed lines) for $M_\chi
	= 0,\dots,7$. $M_\chi$ increases from top to bottom. Blue dotted line marks the state with largest relevant $N_\chi + M_\chi=1+3$ at the curvature $\omega_0/\omega_{\rm c}=0.29$ of the potential valley shown in Fig. 2a of the main text.}%
\label{fig:spectral_weight}%

\end{figure}

The resulting spectral weight is shown in Fig.~\ref{fig:spectral_weight} for
Landau level zero and Landau level one and various values of $M_\chi$.  As
anticipated, Landau level mixing gets stronger for larger values of $N_\chi$
and $M_\chi$. The potential discussed in Fig.~2 of the main text is roughly of
the form $V = \gamma (x^2 + y^2)$ with $\gamma = 0.05\,
\mathrm{meV}/\mathrm{nm}^2$. For $B=14\,\mathrm{T}$ and $m = 0.03\,m_e$, one
obtains from this $\omega_c \approx 54 \, \mathrm{meV}$ and $\omega_0/\omega_c
= \sqrt{2 l_B^2 \gamma/\hbar \omega_c} \approx 0.29$, corresponding to $l_B/L_B
\approx 1.08$.  Eq.~\eqref{eq:radius_avg} bounds the states contributing
notably to the local density of states in our observation range  $r \approx
20\,\mathrm{nm}$ by $N_\chi + M_\chi \leq \langle \vec r^2 \rangle/2 L_B^2 - 1
\approx 4$.  Referring to Fig.~\ref{fig:spectral_weight}, we see that this
implies up to about $5 \%$ Landau level mixing (blue dotted line).

\section{Recursive Green's function method}

For the numerics, we use a tight-binding discretization of the 2D plane with
$N_x$ sites in the $x$-direction, $N_y$ sites in the $y$-direction and
constant lattice spacing $a$. The corresponding Hilbert space is a
tensor-product space of the form $H = H_y \otimes H_x \otimes H_{1/2}$, where
$H_{y/x}$ are $N_{y/x}$-dimensional vector spaces and $H_{1/2}$ is the on-site
Hilbert space of dimension $2$ for a spin-$1/2$ particle.

In order to implement the Rashba Hamiltonian, we need a tight-binding
representation of the covariant derivative $D_j = -i \hbar \partial_{x_j} + e
A_j$ in direction of the unit vector $\vec e_j$. Taylor expansion with respect to $a$
gives the approximate expressions
\begin{align}
\frac{i \hbar}{2a} \biggl( -\psi(\vec x + a \vec e_j)
e^{-i \Phi( \vec x, \vec x+a\vec e_j)}
\nonumber \\ +
\psi(\vec x - a \vec e_j)
e^{i \Phi(\vec x-a\vec e_j, \vec x)}\biggr)
= D_j \psi(\vec x) + \mathcal{O}(a^2), \label{eq:cov_deriv_1}\\
\frac{\hbar^2}{a^2} \biggl( -\psi(\vec x + a \vec e_j)
e^{-i \Phi( \vec x, \vec x+a\vec e_j)} + 2 \psi(\vec x) \nonumber \\
- \psi(\vec x - a \vec e_j)
e^{i \Phi(\vec x-a\vec e_j, \vec x)}\biggr)
= D_j^2 \psi(\vec x) + \mathcal{O}(a^2) \label{eq:cov_deriv_2}
\end{align}
for the covariant derivatives,
where
\begin{align}
\Phi(\vec x_1, \vec x_2) = (-e) \int_{\vec x_1}^{\vec x_2} d \vec x' \cdot \vec A(\vec x')/\hbar
\end{align}
is the Peierls phase accumulated upon going from $\vec x_1$ to $\vec x_2$.
We note that the covariant derivative $D_j \psi(\vec x) =
\langle x | D | \psi\rangle$ in Eq.~\eqref{eq:cov_deriv_1} is evaluated at $\vec
x$, such that terms proportional to $\psi(\vec x + a \vec e_j)$ represent
backwards hopping from $\vec x + a \vec e_j$ to $\vec x$, while the terms
proportional to $\psi(\vec x - a \vec e_j)$ represent forward hopping from
$\vec x - a \vec e_j$ to $\vec x$.

For the tight-binding representation, it is convenient to introduce the
dimensionless hopping matrices $\hat t_{x} \in \mathbb{R}^{N_{x}\times N_{x}}$,
$\hat t_{y} \in \mathbb{R}^{N_y \times N_y}$ which are of the form
\begin{align}
\hat t_{x/y} = \begin{pmatrix}
0 	&  		& 		& 		&  \\
1	& \ddots 	& 	 	& 		&  \\
	& \ddots 	& 		& 		& \\
 	& 		&  1		& 0
\end{pmatrix}
\end{align}
and represent forward hopping from site $i$ to site $i+1$ in the $x/y$ direction.
Using the
Landau gauge $\vec A = -B y \vec e_x$ for a magnetic field $\vec B = \vec
\nabla \times \vec A = B e_z$ in the $z$-direction and the
expressions~\eqref{eq:cov_deriv_1} and \eqref{eq:cov_deriv_2} for the covariant
derivative, one obtains the tight-binding representations
\begin{align}
D_x &\rightarrow \frac{i\hbar}{2a} \sum_{n=1}^{N_y} P_n \otimes
(\hat t_x e^{i \Phi n} - \hat t_x^\dag
e^{-i \Phi n}) \otimes 1 \\
D_y &\rightarrow \frac{i \hbar}{2a}
(\hat t_y - \hat t_y^\dag) \otimes 1 \otimes 1 \\
D_x^2 &\rightarrow \frac{\hbar^2}{a^2}
\sum_{n=1}^{N_y}
P_n \otimes (2 - \hat t_x e^{i \Phi n} - \hat t_x^\dag e^{-i \Phi n}) \otimes 1 \\
D_y^2 &\rightarrow \frac{\hbar^2}{a^2} (2 - \hat t_y - \hat t_y^\dag) \otimes 1 \otimes 1,
\end{align}
of the covariant derivatives. Here, $P_n$ is a projector on site $n$ and $\Phi
= e a^2 B/\hbar = a^2/l_B^2$ is defined such that Peierls phase for hopping in the
$x$-direction can be written as
\begin{align}
\Phi((x_m, y_n), (x_{m+1}, y_n))
= e a B y_n/\hbar = \Phi n,
\end{align}
with $y_n = n a$, $x_m = m a$.
For completeness, we note that $\Phi$ can also be written as $\Phi = \hbar
\omega_c/2 t$ with the hopping energy $t = \hbar^2/2 m a^2$. In total, we find
the tight-binding representation
\begin{align}\label{eq:ham_rashba_tb}
H &=
4 t \, 1 \otimes 1 \otimes 1
-t(\hat t_y + \hat t_y^\dag) \otimes 1 \otimes 1\nonumber \\
& \qquad - \sum_{n=1}^{N_y} t P_n \otimes (e^{i \Phi n} \hat t_x + e^{-i \Phi n} \hat
t_x^\dag
)\otimes 1 \nonumber \\
& \quad
- i \frac{\alpha_R}{2a}(\hat t_y - \hat t_y^\dag) \otimes 1 \otimes \sigma_x \nonumber \\
& \quad + i \frac{\alpha_R}{2a} \sum_{y=1}^{N_y}  P_y \otimes (e^{i \Phi y}
\hat t_x - e^{-i \Phi
	y} \hat t_x^\dag) \otimes \sigma_y \nonumber \\
& \quad + \frac{1}{2} g \mu_B B \, 1 \otimes 1 \otimes \sigma_z \nonumber \\
& \quad
+ \sum_{n=1}^{N_x} \sum_{m=1}^{N_y} V(x=na,y=ma) \, P_m \otimes P_n \otimes 1
\end{align}
of the Rashba-Hamiltonian.
Here, the terms in the first two rows come from the kinetic
energy, the terms in the following two rows are due to the Rashba
spin-orbit interaction, while the terms in the last rows result from the Zeeman
interaction and the disorder potential, respectively.

We use the tight-binding Hamiltonian~\eqref{eq:ham_rashba_tb} to calculate the
single-particle Green's function $G(E, \vec r, \sigma) = \langle \vec
r,\sigma| (E - H + i \eta)^{-1} | \vec r, \sigma\rangle$ at position $\vec r$
and spin-polarization $\sigma$ with the help of the recursive Green's function
algorithm along the $y$-axis~\cite{recursive}.  We use
bilinear spline interpolation to sample the disorder potential on a lattice
with spacing $a = l_B/10$, or equivalently, $a^2/l_B^2 = \hbar \omega_c/2t =
1/100$, such that effects of discretization remain negligble.  In order to
simulate the finite energy resolution within the experiment, we have included a lifetime
broadening of $\eta = 0.05\,\hbar \omega_c$.  The local density of states is
subsequently obtained as $\rho(E, \vec r, \sigma) = -\Im G(E, \vec r,
\sigma)/\pi$.  In order to exclude boundary effects, we sample larger areas
than the area of interest such that the area of interest is at
least $4 l_B$ away from the boundary of the simulation.

We have checked explicitly that our numerics reproduces the local density of
states corresponding to the Fock-Darwin states of a parabolic potential.
As a further cross check, for large samples such as the one shown in Fig.~1a$-$g
of the main text, we calculate the sample-averaged density of states.
As one would expect, for nearly symmetric potentials, the peaks in the
sample-averaged density of states correspond to the
energies~\eqref{eq:rashba_eigenenergies} of the bare Rashba-Hamiltonian.

{\it Note added in proof}: During the referee process of this manuscript, we became aware of another work partly dealing with nodal structure of Landau level wave functions \cite{nematic}.

We acknowledge helpful discussions with S. Florens, T. Champel, F. Hassler, and M. G\"orbig as well as financial support by the German Science Foundation via MO 858/11-2 and INST 222/776-1.
%

\begin{thebibliography}{10}%
\bibitem{Zhang}
M. Z. Hasan and C. L. Kane, Rev. Mod. Phys. {\bf 82}, 3045 (2010);	X. L. Qi and S. C. Zhang, Rev. Mod. Phys. {\bf 83}, 1057 (2011); Y.	Ando,  J. Phys. Soc. Jpn. {\bf 82}, 102001 (2013).
\bibitem{interference}
 C. J.  Davisson  and  L. H.  Germer,  Nature {\bf 119}, 558 (1927); G. P.  Thomson  and  A.  Reid,  Nature  {\bf 119}, 890 (1927);
 C. J\"onsson, Z. Phys. {\bf 161}, 454 (1961); P. G. Merli, G. F. Missiroli, and G. Pozzi, Am. J. Phys. {\bf 44}, 306 (1976); A. Tonomura, J. Endo, T. Matsuda, T. Kawasaki, and H. Ezawa, Am. J. Phys. {\bf 57}, 117 (1989).
\bibitem{Crommie}
M. F. Crommie, C. P. Lutz, and D. M. Eigler, Nature {\bf 363}, 524 (1993); Science {\bf 262}, 218 (1993);
Y. Hasegawa and P. Avouris, Phys. Rev. Lett. {\bf 71}, 1071 (1993);
 J. T. Li,  W. D. Schneider, R. Berndt, and S. Crampin, Phys. Rev. Lett. {\bf 80}, 3332 (1998);
N. Nilius, T. M. Wallis, and W. Ho, Science {\bf 297}, 1853 (2002).
\bibitem{Morgenstern}
M. C. M. M. van der Wielen, A. J. A. van Roij, and H. van Kempen,
Phys. Rev. Lett. {\bf 76}, 1075 (1996);
Chr. Wittneven, R. Dombrowski, M. Morgenstern, and R. Wiesendanger, Phys. Rev. Lett. {\bf 81}, 5616 (1998);
B. Grandidier, Y. M. Niquet, B. Legrand, J. P. Nys, C. Priester, D. Stievenard, J. M. Gerard, and V. Thierry-Mieg, Phys. Rev. Lett. {\bf 85}, 1068 (2000);
T. Maltezopoulos, A. Bolz, C. Meyer, C. Heyn, W. Hansen, M. Morgenstern, and R. Wiesendanger, Phys. Rev. Lett. {\bf 91}, 196804 (2003);
Chr. Meyer, J. Klijn, M. Morgenstern, and R. Wiesendanger, Phys. Rev. Lett. {\bf 91}, 076803 (2003).
\bibitem{Stroscio}
G.M. Rutter, J. N. Crain, N. P. Guisinger,  T. Li, P. N. First, and J. A. Stroscio, Science {\bf 317}, 219 (2007);
Y. B. Zhang, V. W. Brar, C. Girit, A. Zettl, and M. F. Crommie, Nature Phys. {\bf 5}, 722 (2009);
I. Brihuega, P. Mallet, C. Bena, S. Bose, C. Michaelis, L. Vitali, F. Varchon, L. Magaud, K. Kern, and J. Y. Veuillen, Phys. Rev. Lett. {\bf 101}, 206802 (2008);
D. Subramaniam, F. Libisch, Y. Li, C. Pauly, V. Geringer, R. Reiter, T. Mashoff, M. Liebmann, J. Burgdorfer, C. Busse, T. Michely, R. Mazzarello, M. Pratzer, and M. Morgenstern, Phys. Rev. Lett. {\bf 108}, 046801 (2012).
\bibitem{Yazdani}
P. Roushan, J. Seo, C. V. Parker, Y. S. Hor, D. Hsieh, D. Qian, A. Richardella, M. Z. Hasan, R. J. Cava, and A. Yazdani, Nature {\bf 460}, 1106 (2009);
Z. Alpichshev, J. G. Analytis, J. H. Chu, I. R. Fisher, Y. L. Chen, Z. X. Shen, A. Fang, and A. Kapitulnik, Phys. Rev. Lett. {\bf 104}, 016401 (2010);
Y. Okada, C. Dhital, W. Zhou, E. D. Huemiller, H. Lin, S. Basak, A. Bansil, Y.-B. Huang, H. Ding, Z. Wang, S. D. Wilson, and V. Madhavan, Phys. Rev. Lett. {\bf 106}, 206805 (2011).
\bibitem{Schine}
K. Hashimoto, T. Champel, S. Florens, C. Sohrmann, J. Wiebe, Y. Hirayama, R. A. Romer, R. Wiesendanger, and M. Morgenstern, Phys. Rev. Lett. {\bf 109}, 116805 (2012);
N. Schine, A. Ryou, A. Gromov, A. Sommer, and J. Simon, Nature {\bf 534}, 671 (2016).
\bibitem{Landau}
L. D. Landau and E. M. Lifschitz: {\it Quantum Mechanics: Non-relativistic Theory. Course of Theoretical Physics. Vol. 3, 3rd ed.}, Pergamon Press, London (1977).
%
\bibitem{Ando}
 R. Joynt and R. E. Prange, Phys. Rev. B {\bf 29}, 3303 (1984);
 T. Ando, J. Phys. Soc. Jpn. 53, 3101 (1984).
\bibitem{Morgenstern2}
M. Morgenstern, J. Klijn, Chr. Meyer, and R. Wiesendanger, Phys. Rev. Lett. {\bf 90}, 056804 (2003); D. L. Miller,	K. D. Kubista,	G. M. Rutter, M. Ruan,	W. A. de Heer,	M. Kindermann,	P. N. First,	and J. A. Stroscio, Nature Phys. {\bf 6}, 811 (2010).
\bibitem{Hashimoto}
K. Hashimoto, C. Sohrmann, J. Wiebe, T. Inaoka, F. Meier, Y. Hirayama, R. A. Romer, R. Wiesendanger, and M. Morgenstern, Phys. Rev. Lett. {\bf 101}, 256802 (2008).
\bibitem{Hanaguri}
Y.-S. Fu, M. Kawamura, K. Igarashi, H. Takagi, T. Hanaguri, and T. Sasagawa, Nature Phys. {\bf 10}, 815 (2014).
\bibitem{Andrei}
A. Luican-Mayer, M. Kharitonov, G. Li, C.-P. Lu, I. Skachko, A.-M. B. Goncalves, K. Watanabe, T. Taniguchi, and E. Y. Andrei, Phys. Rev. Lett. {\bf 112}, 036804 (2014).
\bibitem{Morgenstern3}
M. Morgenstern, A. Georgi, C. Strasser, C. Ast, S. Becker, and M. Liebmann, Physica E {\bf 44}, 1795 (2012) and references therein.
\bibitem{Bindel}
J. R. Bindel, M. Pezzotta, J. Ulrich, M. Liebmann, E. Sherman, and M. Morgenstern,
Nature Phys. {\bf 12}, 920 (2016).
%
\bibitem{recursive}
P. A. Lee and D. S. Fisher, Phys. Rev. Lett. {\bf 47}, 882 (1981);
D. Thouless and S. Kirkpatrick, J. Phys. C {\bf 14}, 235 (1981);
A. MacKinnon, Z. Phys. B {\bf 59}, 385 (1985).
%
\bibitem{Hernangomez}
T. Champel and S. Florens, Phys. Rev. B {\bf 82}, 045421 (2010);
D. Hernangomez-Perez, J. Ulrich, S. Florens, and T. Champel, Phys. Rev. B {\bf 88}, 245433 (2013);
D. Hernangomez-Perez, S. Florens, and T. Champel, Phys. Rev. B {\bf 89}, 155314 (2014).
\bibitem{Miller}
D. L. Miller, K. D. Kubista, G. M. Rutter, M. Ruan, W. A. de Heer, P. N. First, and J. A. Stroscio, Science {\bf 324}. 924 (2009).
%
\bibitem{supplement}
See supplementary material [http://www.aip.org/pubservs/], which includes Refs. \cite{Feenstra,Dombrowski,Betti,Efros,Becker,Teichmann, Champel,Bychkov,Winkler,Fock,Vogel}.
%
\bibitem{Becker}
S. Becker, M. Liebmann, T. Mashoff, M. Pratzer, and M. Morgenstern, Phys. Rev. B {\bf 81}, 155308 (2010).
%
%
\bibitem{Merkt}
U. Merkt and S. Oelting, Phys. Rev. B {\bf 35}, 2460 (1987).
%
\bibitem{Tsui}
D. C. Tsui, H. L. Stormer, and A. C. Gossard,
Phys. Rev. Lett. {\bf 48}, 1559 (1982).
%
\bibitem{Dean}
K. I. Bolotin,	F. Ghahari,	M. D. Shulman,	H. L. Stormer, and	P. Kim, Nature {\bf 462}, 196 (2009);
X. Du, I. Skachko, F. Duerr, A. Luican, and E. Y. Andrei, Science {\bf 462}, 192 (2009);
 C. R. Dean,	A. F. Young,	P. Cadden-Zimansky,	L. Wang,	H. Ren,	K. Watanabe,	T. Taniguchi,	P. Kim,	J. Hone, and K. L. Shepard, Nature Phys. {\bf 7}, 693 (2011).
%
\bibitem{nematic}
B. E. Feldman, M. T. Randeria, A. Gyenis, F. Wu, H. Ji, R. J. Cava, A. H. MacDonald, and A. Yazdani, Science {\bf 354}, 316 (2016).
\bibitem{Feenstra}
R. M. Feenstra and J. A. Stroscio, J. Vac. Sci. Technol. B {\bf 5}, 923 (1987); M. McEllistrem, G. Haase, D. Chen, and R. J. Hamers,
Phys. Rev. Lett. {\bf 70}, 2471 (1993).
%
\bibitem{Dombrowski}
R. Dombrowski, Chr. Steinebach, Chr. Wittneven, M. Morgenstern, and R. Wiesendanger. Phys. Rev. B {\bf 59}, 8043 (1999).
%
\bibitem{Betti}
M. G. Betti, R. Biagi, U. del Pennino, C. Mariani, and M. Pedio
Phys. Rev. B {\bf 53}, 13605 (1996); M. Getzlaff, M. Morgenstern, Chr. Meyer, R. Brochier, R. L. Johnson, and R. Wiesendanger
Phys. Rev. B {\bf 63}, 205305 (2001).
%
\bibitem{Efros}
A. L. Efros and B I Shklovskii, J. Phys. C {\bf 8}, L49 (1975).
%
\bibitem{Klijn}
M. Morgenstern, J. Klijn, Chr. Meyer, and R. Wiesendanger, Phys. Rev. Lett. {\bf 90}, 056804 (2003).
%
\bibitem{Becker1}
S. Becker, C. Karrasch, T. Mashoff, M. Pratzer, M. Liebmann, V. Meden, and M. Morgenstern
Phys. Rev. Lett. {\bf 106}, 156805 (2011).
%
\bibitem{Teichmann}
J. W. G. Wildoer, A. J. A. van Roij, C. J. P. M. Harmans, and H. van Kempen
Phys. Rev. B {\bf 53}, 10695 (1996); M. Morgenstern, D. Haude, V. Gudmundsson, Chr. Wittneven, R. Dombrowski, and R. Wiesendanger,
Phys. Rev. B {\bf 62}, 7257 (2000); K. Teichmann, M. Wenderoth, S. Loth, R. G. Ulbrich, J. K. Garleff, A. P. Wijnheijmer, and P. M. Koenraad
Phys. Rev. Lett. {\bf 101}, 076103 (2008); F. Marczinowski, J. Wiebe, F. Meier, K. Hashimoto, and R. Wiesendanger
Phys. Rev. B {\bf 77}, 115318 (2008).
%
\bibitem{Champel}
T. Champel and S. Florens, Phys. Rev. B {\bf 80}, 161311(R) (2009); T. Champel, S. Florens, and M. E. Raikh,
Phys. Rev. B {\bf 83}, 125321 (2011).
%
\bibitem{Bychkov}
Y. A. Bychkov and E. I. Rashba,
 JETP Lett. {\bf 39}, 78 (1984).
%
\bibitem{Winkler}
R. Winkler,
 {\em Spin--Orbit Coupling Effects in Two-Dimensional Electron and Hole
  Systems\/}, Vol. 191 of {\em Springer Tracts in Modern Physics\/}
 (Springer, Berlin, Heidelberg, 2003).
%
 \bibitem{Fock}
 V.~Fock, Z. Phys. {\bf 47}, 446 (1928).
%
\bibitem{Vogel}
W. Vogel, D. Welsch, and S. Wallentowitz,
 {\em Quantum Optics: An Introduction\/}
 (Wiley-VCH, Weinheim, 2006).
 %
 %
\end{thebibliography}
\end{document}